\documentclass[12pt]{iopart}
\usepackage{epsfig,graphicx}
\newcommand{\bra}{\langle}
\newcommand{\ket}{\rangle}
\newcommand{\bigbra}{\left\langle}
\newcommand{\bigket}{\right\rangle}
\newcommand{\order}{{\mathcal O}}

\newcommand{\sgn}{\textrm{sgn}}

\newcommand{\one}{{\rm 1\!\!I}}

\newcommand{\be}{\begin{equation}}
\newcommand{\ee}{\end{equation}}
\newcommand{\bd}{\begin{displaymath}}
\newcommand{\ed}{\end{displaymath}}
\newcommand{\vsp}{\vspace*{3mm}}
\newcommand{\room}{\rule[-0.1cm]{0cm}{0.6cm}}

\newcommand{\F}{{\mathcal F}}
\newcommand{\R}{{\rm I\!R}}

\newcommand{\B}{{\mathcal B}}
\newcommand{\D}{{\mathcal D}}
\newcommand{\cP}{{\mathcal P}}

\newcommand{\W}{{\mathcal W}}

\newcommand{\bq}{\ensuremath{\mathbf{q}}}

\newcommand{\bz}{\ensuremath{\mathbf{z}}}

\newcommand{\bR}{\ensuremath{\mathbf{R}}}

\newcommand{\blambda}{{\mbox{\boldmath $\lambda$}}}
\newcommand{\bxi}{{\mbox{\boldmath $\xi$}}}
\newcommand{\bpsi}{{\mbox{\boldmath $\psi$}}}

\newcommand{\bomega}{{\mbox{\boldmath $\omega$}}}
\newcommand{\bsigma}{{\mbox{\boldmath $\sigma$}}}

\newcommand{\bOmega}{{\mbox{\boldmath $\Omega$}}}

\newcommand{\hq}{\hat{q}}

\newcommand{\hA}{\hat{A}}
\newcommand{\hC}{\hat{C}}

\newcommand{\hK}{\hat{K}}
\newcommand{\hL}{\hat{L}}

\newcommand{\here}{\makebox(0,0)}

\newcommand{\rate}{\tilde{\eta}}

\newcommand{\del}{{\delta_{\!N}}}

\newcommand{\notdelta}{\overline{\delta}}
\newcommand{\chiR}{\chi_{\!\!\!~_R}}
\newcommand{\freq}{f}

\begin{document}

\title{Generating functional analysis of Minority Games with real market histories}
\author{A C C Coolen}
\address{
Department of Mathematics, King's College London\\ The Strand,
London WC2R 2LS, UK }

\begin{abstract}
It is shown how the generating functional method of De Dominicis
can be used to solve the dynamics of the original version of the
minority game (MG), in which agents observe real as opposed to
fake market histories. Here one again finds
  exact closed equations for correlation and response
functions, but now these are defined in terms of  two connected
effective non-Markovian stochastic processes: a single effective
agent equation similar to that of the `fake' history models, and a
second effective equation for the overall market bid itself (the
latter is absent in `fake' history models). The result is an exact
theory, from which one can calculate from first principles both
the persistent observables in the MG and  the distribution of
history frequencies.
\end{abstract}


\ead{tcoolen@mth.kcl.ac.uk}

\section{Introduction}

\noindent
 Minority Games (MG) \cite{ChalZhan97} are simple and transparent models
 which were designed to increase our
understanding of the complex collective processes which result
from inductive decision making
 by interacting agents in simplified `markets'. They are
 mathematical implementations of the so-called El Farol bar problem \cite{Arth94}.
 Many versions of
the MG have by now been studied in the literature, see e.g. the
recent textbook  \cite{MGbook} for an overview.
 They differ
in the type of microscopic dynamics used (e.g. batch versus
on-line, stochastic versus deterministic), in the definition of
the information provided to the agents (real-valued versus
discrete, true versus fake market histories) and the agents'
decision making strategies, and also in the specific recipe used
for  converting the observed external information into a trading
action (inner products versus look-up tables).  Models with `fake'
market histories (proposed first in \cite{Cavagna99}), where at
each point in time all agents are given random rather than real
market data upon which to base their decisions, have the advantage
of being Markovian and were therefore the first to be studied and
solved in the theoretical physics literature using techniques from
equilibrium
\cite{ChalMarsZecc00,MarsChalZecc00,MarsChallPRE01,MarsMuletRicciZecch01}
 and
non-equilibrium
\cite{HeimelCoolen01,CoolHeim01,CoolHeimSher01,GallaCoolenSherrington03}
statistical mechanics.

 After \cite{Cavagna99} had revealed the similarity
between the behaviour of the volatility in the standard MG models
with real versus fake market histories, it was shown via numerical
simulations that this statement did not extend to many variations
of the MG, such as games with different strategy valuation update
rules \cite{JohnsonetalPA99b} or with populations where agents do
not all observe history strings of the same length
\cite{JohnsonetalJPA99}. Furthermore, even in the standard MG one
does find profound differences in the history frequency
distributions (although there these differences do not impact on
observables such as the volatility or the fraction of `frozen'
agents).  A partly phenomenological attempt at analyzing
quantitatively the effects of true history in the MG was presented
in \cite{ChalletMarsili00}, and followed by a simulation study
\cite{Lee01} of bid periodicities induced by having real
histories. After these two papers virtually all theorists
restricted themselves to the exclusive analysis of MG versions
with fake histories, simply because  there is no proper theory yet
for MG versions with real histories, in spite of the fact that
these are the more realistic types.

 There would thus seem to be merit in a
 mathematical procedure which would allow for the derivation of
exact dynamical solutions for MGs with {\em real} market
histories. The objective of this paper is to develop and apply
such a procedure. Models with real market histories are strongly
non-Markovian, so analytical approaches based on
pseudo-equilibrium approximations (which require the existence of
a microscopic Lyapunov function) are ruled out. In contrast, the
generating functional analysis (GFA) method of \cite{DeDominicis},
which has an excellent track record in solving the dynamics of
Markovian MGs, will turn out to work also in the case of
non-Markovian models. There are two complications in developing a
GFA for MGs with real histories. Firstly, having real histories
implies that no `batch' version of the dynamics can be defined
(since batch models by definition involve averaging by hand over
all possible histories). Thus one has to return to the original
on-line definitions. Secondly, the temporal regularization method
\cite{BedeLakaShul71} upon which one normally relies in carrying
out a GFA of on-line MG versions is no longer helpful. This
regularization is based on the introduction of random durations of
the individual on-line iteration steps of the process, which
disrupts the timing of all retarded microscopic forces and thereby
leads to extremely messy equations\footnote{Note that in models
with fake histories there are no retarded microscopic forces, so
that there this particular problem could not occur.}. Thus, one
has to develop the GFA directly in terms of the un-regularized
microscopic laws.

This paper is divided into two distinct parts, similar to the more
traditional  GFA studies of MGs with fake market histories. The
first part deals with the derivation of closed macroscopic laws
from which to solve the canonical  dynamic order parameters for
the standard (on-line) MG with true market history. These will
turn out to be formulated in terms of {\em two} effective
equations (rather than a single equation, as for models with fake
histories): one for an effective agent, and one for an effective
overall market bid. These equations are fully exact. The second
part of the paper is devoted to constructing solutions for these
effective processes. In particular, this paper focuses on the
usual persistent observables of the MG and on the distribution of
history frequencies, which are calculated in the form of an
expansion of which the first few terms are derived in explicit
form. The final results find excellent confirmation in numerical
simulations.

\section{Definitions}

\subsection{Generalized Minority Game with both valuation and overall
bid perturbations}

In the standard MG one imagines having $N$ agents, labeled by
$i=1,\ldots,N$. At each iteration step $\ell\in\{0,1,2,\ldots\}$
of the game, each agent $i$ submits a `bid' $b_i(\ell)\in\{-1,1\}$
to the market. The (re-scaled) cumulative market bid at stage
$\ell$ is defined as
\begin{equation} A(\ell)=\frac{1}{\sqrt{N}}\sum_{i=1}^N b_i(\ell)+A_e(\ell)
\label{eq:define_bid}
\end{equation}
An external contribution $A_e(\ell)$ has been added, representing
e.g. the actions of market regulators, which will enable us to
identify specific response functions later. Profit is assumed to
be made by those agents who find themselves subsequently in the
minority group, i.e. when $A(\ell)>0$ by those agents $i$ with
$b_i(\ell)<0$, and when $A(\ell)<0$ by those with $b_i(\ell)>0$.
 Each agent $i$ determines his
bid $b_i(\ell)$ at each step $\ell$ on the basis of publicly
available information, which the agents believe to represent
historic market data, here given by the vector
$\blambda(\ell,A,Z)\in\{-1,1\}^M$:
\begin{equation}
\blambda(\ell,A,Z)=\left(\begin{array}{c}
\sgn\big[(1-\zeta)A(\ell-1)+\zeta Z(\ell,1)\big]\\
 \vdots
\\
\sgn\big[(1-\zeta)A(\ell-M)+\zeta Z(\ell,M)\big]
\end{array}\right)
\label{eq:define_info_LU}
\end{equation}
The numbers $\{Z(\ell,\lambda)\}$, with $\lambda=1,\ldots,M$, are
zero-average Gaussian random variables, which represent a `fake'
alternative to the true market data.  $M$ is the number of
iteration steps in the past for which market information is made
available. We define $\alpha=2^M/N$, and take $\alpha$ to remain
finite as $N\to\infty$. The parameter $\zeta\in[0,1]$ allows us to
interpolate between the cases of strictly true ($\zeta=0$) and
strictly fake ($\zeta=1$) market histories. We
 distinguish between two classes of `fake history' variables:
\begin{eqnarray}
{\rm consistent:}~ & Z(\ell,\lambda)=Z(\ell-\lambda),~ & \bra
Z(\ell)Z(\ell^\prime)\ket=\kappa^2\delta_{\ell \ell^\prime}~~~~
\label{eq:typeA}
\\
{\rm inconsistent:}~ & Z(\ell,\lambda)~{\rm all~independent},~ &
\bra
Z(\ell,\lambda)Z(\ell^\prime\!,\lambda^\prime)\ket=\kappa^2\delta_{\ell
\ell^\prime}\delta_{\lambda\lambda^\prime}~~~~ \label{eq:typeB}
\end{eqnarray}
We note that (\ref{eq:typeB}) does not correspond to a pattern
being shifted in time, contrary to what one expects of a string
representing the time series of the overall bid, so that the
agents in a real market could easily detect that they are being
fooled. Hence (\ref{eq:typeA}) seems a more natural description of
fake history. Although fake, it is at least consistently so.

Each agent has $S$ trading strategies, which we label by
$a=1,\ldots,S$. Each strategy $a$ of each trader $i$ consists of a
complete list $\bR^{ia}$ of $2^M$ recommended  trading decisions
$\{R_{\blambda}^{ia}\}\in\{-1,1\}$, covering all $2^M$ possible
values of the external information vector $\blambda$. We draw all
entries $\{R^{ia}_{\blambda}\}$ randomly and independently
 before the start of the game, with equal probabilities for $\pm 1$.
  Upon observing history string $\blambda(\ell,A,Z)$
at stage $\ell$, given a trader's active strategy at that stage is
$a_{i}(\ell)$, the agent  will follow the instruction of his
active strategy and take the decision $b_i(\ell)= R^{i
a_i(\ell)}_{\blambda(\ell,A,Z)}$. To determine their active
strategies\index{active strategy} $a_i(\ell)$, all agents keep
track of valuations $p_{ia}(\ell)$, which measure how often and to
what extent each strategy $a$ would have led to a minority
decision if it had been used from the start of the game onwards.
These valuations are updated continually, via
 \begin{equation}
 p_{ia}(\ell+1)=p_{ia}(\ell)-\frac{\rate}{\sqrt{N}}A(\ell)R^{ia}_{\blambda(\ell,A,Z)}
\end{equation}
The factor $\rate$ represents a learning rate. If the active
strategy $a_{i}(\ell)$ of trader $i$ at stage $\ell$ is defined as
 the one with the highest valuation $p_{ia}(\ell)$ at that point, and upon writing $\F_{\blambda}[\ell,A,Z]=\sqrt{\alpha
N}~\delta_{\blambda,\blambda(\ell,A,Z)}$, our process becomes
\begin{eqnarray}
p_{ia}(\ell+1)&= &p_{ia}(\ell)-\frac{\rate}{N\sqrt{\alpha
}}A(\ell)\sum_{\blambda} R_{\blambda}^{ia}\F_\blambda[\ell,A,Z]
  \label{eq:valuation_dynamics}
\\ A(\ell)&=&A_e(\ell)+ \frac{1}{N\sqrt{\alpha}}\sum_{i}
\sum_{\blambda} R_{\blambda}^{i a_i(\ell)} \F_\blambda[\ell,A,Z]
 \label{eq:totalbid_eqn}
 \\
a_{i}(\ell)&=& {\rm arg}~\max_{a\in\{1,\ldots,S\}}
\{p_{ia}(\ell)\}
\end{eqnarray}
We note that $(\alpha N)^{-1}\sum_{\blambda}1=(\alpha
N)^{-1}\sum_{\blambda}\F_\blambda^2[\ell,A,Z]=1$.  The standard
 MG  is recovered for $\zeta\to 0$
(i.e. true market data only), whereas the `fake history'  MG  as
in e.g. \cite{Cavagna99,HeimelCoolen01} is found for $\zeta\to 1$
(i.e. fake market data only, of the inconsistent type
(\ref{eq:typeB})).

Henceforth we will restrict ourselves to the simplest case $S=2$,
where each agent has only two strategies, so $a\in\{1,2\}$, since
 the choice
made for $S$ has been shown to have  only a quantitative effect on
the behaviour of the MG. Our equations can now be simplified in
the standard way upon introducing the new variables
\begin{eqnarray}
q_i(\ell)=\frac{1}{2}[p_{i1}(\ell)- p_{i2}(\ell)]~~~~~~~~~~~~~~~
\label{eq:define_q} \\
\bomega^i=\frac{1}{2}[\bR^{i1}+\bR^{i2}],~~~~~~~
\bxi^i=\frac{1}{2}[\bR^{i1}-\bR^{i2}] \label{eq:define_xi}
\end{eqnarray}
 and
$\bOmega=N^{-1/2}\sum_i \bomega^i$.  The bid of agent $i$ at step
$\ell$ is now seen to follow from
\begin{eqnarray}
\bR^{i a_i(\ell)}
 &=&\frac{1}{2}[\bR^{i 1} -\bR^{i 2}
] +\frac{1}{2}\sgn[q_i(\ell)][\bR^{i 1} +\bR^{i 2} ]\nonumber\\
&=& \bomega^i+\sgn[q_i(\ell)]\bxi^i
\end{eqnarray}
 The above $S=2$ formulation is
easily generalized to include decision noise: one simply replaces
$\sgn[q_i(\ell)]\to \sigma[q_i(\ell),z_i(\ell)]$, in which the
$\{z_{j}(\ell)\}$ are independent and zero average random numbers,
described by a symmetric and unit-variance distribution $P(z)$.
The function $\sigma[q,z]$ is taken to be
 non-decreasing in $q$ for any $z$, and
 parametrized by a control parameter $T\geq 0$
 such that $\sigma[q,z]\in \{-1,1\}$, with $\lim_{T\to
 0}\sigma[q,z]=\sgn[q]$ and
 $\lim_{T\to\infty}\int\!dz~P(z)\sigma[q,z]=0$.
Typical examples are
 additive and multiplicative noise definitions, described by
$\sigma[q,z]=\sgn[q+Tz]$ and
  $\sigma[q,z]=\sgn[q]~\sgn[1+Tz]$, respectively.
The parameter
 $T$ measures
  the degree of randomness in the agents' decision making,
  with $T=0$ bringing us back to  $a_{i}(\ell)={\rm arg}~\max_a \{p_{ia}(\ell)\}$,
  and with purely random strategy selection for $T=\infty$.

Upon translating our microscopic laws
(\ref{eq:valuation_dynamics},\ref{eq:totalbid_eqn}) into the
language of the valuation differences (\ref{eq:define_q})
 for $S=2$, we find that now our MG equations close in terms of our new dynamical
  variables $\{q_i(\ell)\}$, so that perturbations of valuations (again for the purpose of defining
response functions later) can be implemented simply by replacing
$q_i(\ell)\to q_{i}(\ell)+\theta_i(\ell)$, with
$\theta_i(\ell)\in\R$. Thus we arrive at the following closed
equations,  defining our generalized $S=2$ MG process:
\begin{eqnarray}
&& q_{i}(\ell+1)
=
q_{i}(\ell)+\theta_i(\ell)-\frac{\rate}{N\sqrt{\alpha
}}\sum_{\blambda} \xi_{\blambda}^{i}\F_\blambda[\ell,A,Z]
 A(\ell)
 \label{eq:mem_qeqn}
\\
&& A(\ell)= A_e(\ell)+\frac{1}{\sqrt{\alpha N}}
\sum_{\blambda}\Big\{\Omega_{\blambda}\!+\frac{1}{\sqrt{N}}\sum_i\sigma[q_i(\ell),z_i(\ell)]\xi_\blambda^i\Big\}
\F_\blambda[\ell,A,Z] ~~~~~~~~\label{eq:mem_Aeqn}
\\
&& \F_{\blambda}[\ell,A,Z]=\sqrt{\alpha
N}~\delta_{\blambda,\blambda(\ell,A,Z)} \label{eq:mem_Feqn}
\\
&& \blambda(\ell,A,Z)=\left(\begin{array}{c}
\sgn[(1-\zeta)A(\ell-1)+\zeta Z(\ell,1)]\\
 \vdots
\\
\sgn[(1-\zeta)A(\ell-M)+\zeta Z(\ell,M)]
\end{array}\right)
\label{eq:mem_lambda}
\end{eqnarray}
 The values of $\{A(\ell),Z(\ell)\}$ for $\ell\leq 0$ and of the $q_i(0)$
play the role of initial conditions.

The key differences at the mathematical level between MG models
with fake history and those with true history as defined above,
are in the dependence of the microscopic laws on the past via the
history string $\{A(\ell-1),\ldots,A(\ell-M)\}$ occurring in
$\blambda(\ell,A,Z)\in\{-1,1\}^M$, in combination with the
presence and role of the zero-average Gaussian random variables
$\{Z(\ell,\lambda)\}$.

\subsection{Mathematical consequences of having real history}

In all generating functional analyses of MGs which have been
published so far, the choice $\zeta=1$ eliminated with one stroke
of the pen the dependence of the process on the history
$\{A(\ell-1),\ldots,A(\ell-M)\}$.  The variables
$\{Z(\ell,1),\ldots,Z(\ell,M)\}$ could subsequently be replaced
simply by integer numbers $\mu$, labeling each of the
$2^M=p=\alpha N$ possible `pseudo-histories' that could have been
drawn at any given time step $\ell$. Here this is no longer
possible. The variables $\{Z(\ell,\lambda)\}$ now play the role of
random disturbances of the true market history as perceived by the
agents, and there is no reason why all possible histories should
occur (let alone with equal frequencies) or why some entries
$\{Z(\ell,\lambda)\}$ (e.g. those with small values of $\lambda$,
which  corrupt the most recent past in the history string) could
not be more important than others. The problem  has become {\em
qualitatively} different. One can thus anticipate various
mathematical consequences for the generating functional analysis
of introducing history into the MG. An early appreciation of these
will help us to proceed with the calculation more efficiently.

Firstly, we will have to analyze the original on-line version of
the MG; the batch version can no longer exist by definition, since
it involves averaging by hand over all possible `histories' at
each iteration step.  However, the temporal regularization method
of \cite{BedeLakaShul71} which was employed successfully for the
on-line MG with fake history \cite{CoolHeim01}, based on
introducing Poissonnian distributed real-valued random durations
for the individual iterations in
(\ref{eq:mem_qeqn},\ref{eq:mem_Aeqn}), can in practice no longer
be used in the non-Markovian case. The reason for this is the
problem which prompted the authors of \cite{CoolHeim01} to add the
external perturbations $\theta_i(\ell)$ to the regularized on-line
process rather than
 to the original equations: whereas in a
Markov chain the introduction of random durations for the
individual iteration steps only implies a harmless uncertainty in
where we are on the time axis, in a system with retarded
interactions one would generate very messy equations. We must
therefore proceed with our process as it is, without temporal
regularization (although we will be able to recover the previous
theory in the limit $\zeta\to 1$, as it should). It will in fact
turn out that the more direct application of the generating
functional method presented in this paper brings the benefit of
greater transparency. For instance, the continuity assumptions
underlying our use of saddle-point arguments in path integrals
become much more clear than they were in \cite{CoolHeim01}. As
always we continue to concentrate on the evaluation and disorder
averaging of the generating functional
\begin{equation}
 Z[\bpsi]=\bra
e^{i\sum_{\ell>0}\sum_i
\psi_i(\ell)\sigma[q_i(\ell),z_i(\ell)]}\ket
 \label{eq:memZ}
\end{equation}
 The
brackets in (\ref{eq:memZ}) denote averaging over the stochastic
process (\ref{eq:mem_qeqn},\ref{eq:mem_Aeqn}), whose randomness is
here caused by the decision noise $\{\bz(\ell)\}$ and the fake
history variables $\{Z(\ell,\lambda)\}$. Although  (\ref{eq:memZ})
looks like the corresponding expressions
 for batch MGs, here we have to
allow for $\ell=\order(N)$. Studying the un-regularized process
also implies that one has to be more careful with finite size
corrections. This has consequences in working out the disorder
average of the generating functional: in previous MG versions one
needed only the first two moments of the distribution of the
strategy look-up table entries. Here, although one must still
expect only the first two moments to play a role in the final
theory, the need to keep track initially of the finite size
correction terms implies that our equations simplify considerably
if, instead of binary strategy entries, we choose the variables
$\{R_{\blambda}^{ia}\}$ to be zero-average and unit-variance
Gaussian variables.

 It will turn out that in our
analysis of (\ref{eq:memZ}) an important role will be played by
the following quantity:
\begin{eqnarray}
\overline{W}[\ell,\ell^\prime;A,Z]&=& \frac{1}{\alpha
N}\sum_{\blambda}\F[\ell,A,Z]\F[\ell^\prime,A,Z]\nonumber
\\
&=& \delta_{\blambda(\ell,A,Z),\blambda(\ell^\prime,A,Z)}
\label{eq:overlineW}
\end{eqnarray}
This object is a function of the paths $\{A\}$ and $\{Z\}$, and
indicates whether or not the histories {\em as perceived by the
agents} at times $\ell$ and $\ell^\prime$ are identical
(irrespective of the extent to which these `histories' are true).
Its statistics are trivial in the absence of history, but will
here  generally contain information regarding the recurrence of
overall bid trajectories. For reasons of economy we will formulate
our theory in terms of the quantity (\ref{eq:overlineW}), rather
than substitute
$\delta_{\blambda(\ell,A,Z),\blambda(\ell^\prime,A,Z)}$ directly.
This will prevent unnecessary future repetition, since it will
allow for most of the theory to be applied also to MG models with
inner product rather than look-up table definitons for the agents'
history-to-action conversion \cite{InoueCoolen}.

\section{The disorder averaged generating functional}

\subsection{Evaluation of the disorder average}

Rather than first writing the microscopic process in probabilistic
form, as in \cite{CoolHeim01}, we will  express the generating
functional (\ref{eq:memZ}) as an integral over all possible joint
paths of the state vector $\bq$ and of the overall bid $A$, and
insert appropriate $\delta$-distributions to enforce the
microscopic dynamical equations
(\ref{eq:mem_qeqn},\ref{eq:mem_Aeqn}), i.e.
\begin{eqnarray*}
1&=&\prod_{i\ell}\int\!\left[\frac{d\hq_i(\ell)}{2\pi}\right]e^{i\hq_i(\ell)[
q_{i}(\ell+1)-q_{i}(\ell)-\theta_i(\ell)+\frac{\rate}{N\sqrt{\alpha
}}\sum_{\blambda} \xi_{\blambda}^{i}\F_\blambda[\ell,A,Z] ]
A(\ell)}\\
1&=&\prod_{\ell}\int\!\left[\frac{d\hA(\ell)}{2\pi}\right]e^{i\hA(\ell)[
A(\ell)-A_e(\ell)-\frac{1}{\sqrt{\alpha N}}
\sum_{\blambda}\Big\{\Omega_{\blambda}+\frac{1}{\sqrt{N}}\sum_i\sigma[q_i(\ell),z_i(\ell)]\xi_\blambda^i\Big\}
\F_\blambda[\ell,A,Z] }
\end{eqnarray*}
(since our microscopic laws are of an iterative and causal form,
they have unique solutions). To compactify our equations we will
use the short-hand $s_i(\ell)=\sigma[q_i(\ell),z_i(\ell)]$. We can
now write the disorder average $\overline{Z[\bpsi]}$ of
(\ref{eq:memZ}) as follows:
 \begin{eqnarray}
\overline{
Z[\bpsi]}&=&\int\!\left[\prod_{\ell>0}\frac{dA(\ell)d\hA(\ell)
}{2\pi}e^{i\hA(\ell)[A(\ell)-A_e(\ell)]}\right]
 \nonumber
\\
&&\times\bigbra
\int\!\left[\prod_{i\ell}\frac{dq_i(\ell)d\hq_i(\ell)}{2\pi}e^{i\hq_i(\ell)[
q_{i}(\ell+1)-q_{i}(\ell)-\theta_i(\ell)]+i\psi_i(\ell)s_i(\ell)}
\right] \right. \label{eq:memZbeforescaling}
\\
&& \left.\times~\overline{ e^{\frac{i}{N\sqrt{\alpha
}}\sum_{\blambda}\sum_{i\ell}\left[
 \rate\hq_i(\ell) \xi_{\blambda}^{i} A(\ell)-\hA(\ell)
\Big(\omega^i_{\blambda}+ s_i(\ell)\xi_\blambda^i\Big) \right]
\F_\blambda[\ell,A,Z]}}~\bigket_{\!\{\bz,Z\}} \nonumber
\end{eqnarray}
The brackets $\bra \ldots\ket_{\{\bz,Z\}}$ denote averaging over
the Gaussian decision noise and the pseudo-memory variables, and
we have used the abbreviations (\ref{eq:define_xi}).
 The short-hand $Du=(2\pi)^{-\frac{1}{2}}e^{-\frac{1}{2}u^2}$
 and the previously introduced quantity $\overline{W}[\ldots]$
in (\ref{eq:overlineW})
 allow us to
 write
the disorder average (over the independently distributed
zero-average and unit-variance $R_{\blambda}^{ia}$) in the last
line of (\ref{eq:memZbeforescaling}) as
\begin{eqnarray}
\overline{e^{\frac{i}{N\sqrt{\alpha}}\sum_{\blambda}^{~}\!\sum_{i\ell}\ldots}}
&=& \prod_{\blambda}\prod_i \int\!\!Du~
e^{\frac{iu}{2N\sqrt{\alpha }}\sum_{\ell}\left[
 \rate\hq_i(\ell) A(\ell)-\hA(\ell)
[1+s_i(\ell)] \right] \F_\blambda[\ell,A,Z]} \nonumber
\\
&& \times\prod_{\blambda}\prod_i \int\!\!Dv~
e^{\frac{iv}{2N\sqrt{\alpha }}\sum_{\ell}\left[
 \rate\hq_i(\ell) A(\ell)+\hA(\ell)
[1-s_i(\ell)] \right] \F_\blambda[\ell,A,Z]} \nonumber
\\
&&\hspace*{-15mm} =
e^{-\frac{1}{4N}\sum_{\ell\ell^\prime>0}\overline{W}[\ell,\ell^\prime;A,Z]
\sum_{i}\left[
 \rate\hq_i(\ell) A(\ell)-\hA(\ell)s_i(\ell) \right]
\left[
 \rate\hq_i(\ell^\prime) A(\ell^\prime)-\hA(\ell^\prime)s_i(\ell^\prime) \right] }
 \nonumber
\\
&&\times~\room
e^{-\frac{1}{4}\sum_{\ell\ell^\prime>0}\hA(\ell)\overline{W}[\ell,\ell^\prime;A,Z]
\hA(\ell^\prime)} \label{eq:mem_disav}
\end{eqnarray}
 We next  isolate the usual
observables
$L(\ell,\ell^\prime)=N^{-1}\sum_i\hq_i(\ell)\hq_i(\ell^\prime)$,
$K(\ell,\ell^\prime)=N^{-1}\sum_i s_i(\ell)\hq_i(\ell^\prime)$,
and $C(\ell,\ell^\prime)=N^{-1}\sum_i s_i(\ell) s_i(\ell^\prime)$,
by inserting appropriate integrals over $\delta$-distributions. We
also use the abbreviations $\D C=\prod_{\ell
\ell^\prime}[\sqrt{N}dC(\ell,\ell^\prime)/\sqrt{2\pi}]$ (similarly
for other two-time observables) and  $\D A=\prod_{\ell>0}
[dA(\ell)/\sqrt{2\pi}]$ (similarly for $\hat{A}$). Initial
conditions for the $q_i(0)$ are assumed to be of the factorized
form $p_0(\bq)=\prod_i p_0(q_i(0))$. In anticipation of issues to
arise in  subsequent stages of our analysis, especially those
related to the scaling with $N$ of the number of individual
iterations of the process, we will also define the largest
iteration step in the generating functional as $\ell_{\rm max}$.
All this allows us to write $\overline{Z[\bpsi]}$ in the form
 \begin{eqnarray}
\overline{ Z[\bpsi]}&=&\int\! \D C\D \hC\D K\D \hK\D L\D
\hL~e^{iN\sum_{\ell\ell^\prime}[\hC(\ell,\ell^\prime)C(\ell,\ell^\prime)+\hK(\ell,\ell^\prime)K(\ell,\ell^\prime)+\hL(\ell,\ell^\prime)
L(\ell,\ell^\prime)]} \nonumber \\ &&\times ~ e^{\order(\ell_{\rm
max}^2\log N) } \int\!\D A\D \hA~e^{i\sum_\ell
\hA(\ell)[A(\ell)-A_e(\ell)]} \nonumber
\\
&& \room \times~
e^{\frac{1}{4}\rate\sum_{\ell\ell^\prime}\overline{W}[\ell,\ell^\prime;A,Z]\left\{
\hA(\ell)K(\ell,\ell^\prime)A(\ell^\prime)
+\hat{A}(\ell^\prime)K(\ell^\prime,\ell)A(\ell)
 \right\}}
\nonumber
\\
&& \room \times ~
e^{-\frac{1}{4}\sum_{\ell\ell^\prime}\overline{W}[\ell,\ell^\prime;A,Z]\left\{
\rate^2 A(\ell)L(\ell,\ell^\prime)A(\ell^\prime)
+\hA(\ell)[1+C(\ell,\ell^\prime)]\hA(\ell^\prime)
 \right\}}
 \nonumber
\\
&& \times
 \bigbra\int\!\prod_{i
\ell}\left[\frac{dq_i(\ell)d\hq_i(\ell)}{2\pi} e^{i\hq_i(\ell)[
q_{i}(\ell+1)-q_{i}(\ell)-\theta_i(\ell)]+ i\psi_i(\ell)s_i(\ell)}
\right].\prod_i p_0(q_i(0)) \right.\nonumber \\ &&
\left.\hspace*{5mm} \times \prod_i e^{-i\sum_{\ell\ell^\prime}\{
\hL(\ell,\ell^\prime)\hq_i(\ell)\hq_i(\ell^\prime)+\hK(\ell,\ell^\prime)s_i(\ell)\hq_i(\ell^\prime)
+\hC(\ell,\ell^\prime)s_i(\ell)s_i(\ell^\prime)\}}
\bigket_{\!\{\bz,Z\}}
 \nonumber
  \\ &=& \int\!\D C\D \hC\D K\D \hK\D
L\D \hL~e^{N[\Psi+\Omega+\Phi]+\order(\ell_{\rm max}^2\log N)}
\label{eq:truesaddlepoint}
\end{eqnarray}
with
\begin{eqnarray}
\Psi &=&i\sum_{\ell\ell^\prime\leq \ell_{\rm max}
}\big[\hC(\ell,\ell^\prime)C(\ell,\ell^\prime)+\hK(\ell,\ell^\prime)K(\ell,\ell^\prime)+\hL(\ell,\ell^\prime)L(\ell,\ell^\prime)
\big] \label{eq:truePsi}
\\
\Phi&=&\frac{1}{N}\log
 \bigbra\int\!\!\D A\D \hA~
e^{i\sum_{\ell\leq \ell_{\rm max}} \hA(\ell)[A(\ell)-A_e(\ell)]}
\nonumber\right.
\\
&&\left.\hspace*{20mm}\times~
 e^{-\frac{1}{4}\sum_{\ell\ell^\prime\leq
\ell_{\rm max}
}\overline{W}[\ell,\ell^\prime;A,Z]M[\ell,\ell^\prime;A,\hA]}
 \bigket_{\!\{Z\}}
\label{eq:truePhi}\\
 \Omega&=& \frac{1}{N}\sum_i \log
\bigbra\int\!\left[\prod_{\ell=0}^{\ell_{\rm max}
}\frac{dq(\ell)d\hq(\ell)}{2\pi}\right] p_0(q(0)) \right.
\nonumber
 \\ &&
  \left.\room \times~ e^{i\sum_{\ell\leq \ell_{\rm
max}}\!\big[ \hq(\ell)[ q(\ell+1)-q(\ell)-\theta_i(\ell)]+
\psi_i(\ell)\sigma[q(\ell),z(\ell)]\big]
-i\sum_{\ell\ell^\prime\leq \ell_{\rm max}}\!
\hq(\ell)\hL(\ell,\ell^\prime)\hq(\ell^\prime) }\right. \nonumber
\\ && \left.\room \times~e^{-i\sum_{\ell\ell^\prime\leq
\ell_{\rm max}}\big[
\hC(\ell,\ell^\prime)\sigma[q(\ell),z(\ell)]\sigma[q(\ell^\prime),z(\ell^\prime)]
+ \hK(\ell,\ell^\prime)\sigma[q(\ell),z(\ell)]\hq(\ell^\prime)
\big]} \bigket_{\!\bz}
 \label{eq:trueOmega}
\end{eqnarray}
and with
\begin{eqnarray}
 M[\ell,\ell^\prime;A,\hA]&=&
\rate^2 A(\ell)L(\ell,\ell^\prime)A(\ell^\prime) -
\rate\big[\hA(\ell)K(\ell,\ell^\prime)A(\ell^\prime) +
\hA(\ell^\prime)K(\ell^\prime\!,\ell)A(\ell)\big] \nonumber
\\[1mm]
&& +\hA(\ell)[1+C(\ell,\ell^\prime)]\hA(\ell^\prime)
 \end{eqnarray}
 The $\order(\ell_{\rm max}^2\log N)$ corrections in
(\ref{eq:truesaddlepoint}) are constants, which reflect
  the scaling with $N$ used in defining the conjugate order
 parameters.

Compared to the Markovian (fake history) MG versions, we note that
$\Psi$ and $\Omega$ take their conventional forms, and that all
the complications induced by having true market history are
concentrated in the function $\Phi[C,K,L]$, which is now defined
in terms of a stochastic process for the overall bid $A(\ell)$
rather than being an explicit function of the order parameters
(which had been the situation in all fake history versions of the
game), and in the remaining task to implement an appropriate
scaling with $N$ of the time scale $\ell_{\rm max}$. We can now
also see the advantage in our earlier decision to define Gaussian
rather than binary  look-up table entries. With the $N$-scaling of
$\ell_{\rm max}$ still pending, instead of (\ref{eq:mem_disav}),
in the binary case we  would have found
\begin{eqnarray}
\overline{e^{\frac{i}{N\sqrt{\alpha}}\sum_{\blambda}^{~}\!\sum_{i\ell}\ldots}}
&=& e^{\sum_{i}\sum_{\blambda}\log\cos\Big[\frac{1}{2N\sqrt{\alpha
}}\sum_{\ell\leq \ell_{\rm max}}\left[
 \rate\hq_i(\ell) A(\ell)-\hA(\ell)
[1+s_i(\ell)] \right] \F_\blambda[\ell,A,Z]\Big]} \nonumber
\\
&& \hspace*{-13mm} \times~ e^{\sum_{i}\sum_{\blambda}\log
\cos\Big[\frac{1}{2N\sqrt{\alpha }}\sum_{\ell\leq \ell_{\rm max}}
\left[
 \rate\hq_i(\ell) A(\ell)+\hA(\ell)
[1-s_i(\ell)] \right] \F_\blambda[\ell,A,Z]\Big]}~~~~~
\label{eq:mem_disav_com}
\end{eqnarray}
In this expression we see that, for $\ell_{\rm max}=\order(N)$,
the different choices of strategy look-up table entry distribution
will give the same results only for those paths $\{A,Z\}$ where
the frequency of occurrence each of the $2^M$ possible histories
is of order $\order(N^{-1})$. In the latter case the function
$\F_\blambda[\ell,A,Z]$ scales effectively inside summations over
$\ell$ as $\F_\blambda[\ell,A,Z]=\order(N^{-\frac{1}{2}})$, and we
return to (\ref{eq:mem_disav}). Thus, for non-Gaussian
distributions of the $\{R_{\blambda}^{ia}\}$ at this stage of the
GFA one either has to carry on with the more complicated
expression (\ref{eq:mem_disav_com}), which cannot be expressed in
terms of the order parameters $\{C,K,L\}$, or one has to make
further assumptions on the overall bid statistics, which (although
turning out to be correct) require validation {\em a posteriori}.

\subsection{Canonical time scaling}

For the on-line MG with random external information (i.e. with
$\zeta=1$) it is known that the relevant time scale is $\ell_{\rm
max}=\order(N)$. Rather than imposing the time scale $\ell_{\rm
max}=\order(N)$ by hand, it is satisfactory to see that one can
also extract this canonical time scaling from our present
equations
(\ref{eq:truesaddlepoint},\ref{eq:truePsi},\ref{eq:truePhi},\ref{eq:trueOmega}).

 For finite $\ell_{\rm max}$ we immediately find
$\lim_{N\to\infty}\Phi=0$ in (\ref{eq:truePhi}), and our
generating functional will be dominated by the physical
saddle-point of $\lim_{N\to\infty}[\Psi+\Omega]$, giving
$\hC=\hK=\hL=0$. This leads to a trivial effective single spin
problem, which just  describes a frozen state. This makes perfect
sense in view of our definitions
(\ref{eq:mem_qeqn},\ref{eq:mem_Aeqn}): individual updates of the
variables $q_i$ are of order $N^{-\frac{1}{2}}$, so nothing can
change on time-scales corresponding to only a finite number of
iteration steps.  Thus our present equations automatically lead us
to the choice $\ell_{\rm max}=\order(1/\del)$, where
$\lim_{N\to\infty}\del=0$; the function $\Phi$ will indeed scale
differently as soon as $\ell_{\rm max}$ is allowed to diverge with
$N$.
 We thus define $\ell_{\rm max}=t_{\rm max}/\del$, where $0\leq
t_{\rm max}<\infty$ (of order $N^0$) and with
$\lim_{N\to\infty}\del=0$. In order to obtain well-defined limits
at the end in (\ref{eq:truePsi}), we see that we have to re-scale
our conjugate order parameters according to $(\hC,\hK,\hL)\to
\del^2 (\hC,\hK,\hL)$. Furthermore, for the perturbation fields
$\{\theta_i,\psi_i\}$ to retain statistical significance they also
will have to be re-scaled in the familiar manner, according to
$(\theta_i,\psi_i)\to
\del^{\!\!-1}(\tilde{\theta}_i,\tilde{\psi}_i)$ (similar to
\cite{CoolHeim01}). The integrations over order parameters and
conjugate order parameters in (\ref{eq:truesaddlepoint}) will now
become path integrals for $N\to\infty$\footnote{This is the point,
therefore, where the inevitable continuity assumptions  regarding
our macroscopic dynamic observables enter. In the present
derivation these
 take a more transparent form than in \cite{CoolHeim01},
where they were hidden inside the details of the temporal
regularization.}.

 It will be convenient to introduce the following effective
measure:
\begin{eqnarray}
\hspace*{-15mm}
 \bra
g[\{q,\hq,z\}]\ket_\star&=&\lim_{N\to\infty}\frac{1}{N}\sum_i
\frac{ \int\!\prod_{\ell=1}^{t_{\rm max}
/\del}[dq(\ell)d\hq(\ell)]\bra M_i[\{q,\hq,z\}]
g[\{q,\hq,z\}]\ket_{\bz}}{ \int\!\prod_{\ell=1}^{t_{\rm max}
/\del}[dq(\ell)d\hq(\ell)]\bra M_i[\{q,\hq,z\}]\ket_{\bz}}
\\
\hspace*{-15mm} \room M_i[\{q,\hq,z\}]&=& p_0(q(0))
e^{i\delta_N\sum_{\ell=1}^{t_{\rm max} /\del}\hq(\ell)\Big[
\frac{q(\ell+1)-q(\ell)}{\del}-\tilde{\theta}_i(\ell)\Big]+
i\del\sum_\ell\tilde{\psi}_i(\ell)\sigma[q(\ell),z(\ell)]}
\nonumber\\
 &&
  \hspace*{-15mm}
  \room
  \times~e^{
-i\del^2\!\sum_{\ell\ell^\prime=1}^{t_{\rm max}/\del}\Big[
\hL(\ell,\ell^\prime)\hq(\ell)\hq(\ell^\prime)
+\hK(\ell,\ell^\prime)\sigma[q(\ell),z(\ell)]\hq(\ell^\prime)+\hC(\ell,\ell^\prime)\sigma[q(\ell),z(\ell)]
\sigma[q(\ell^\prime),z(\ell^\prime)]\Big] }\nonumber
\\
&& \label{eq:measureMi}
\end{eqnarray}
Upon substituting $\ell_{\rm max}=t_{\rm max}/\del$ into our
equations (\ref{eq:truePsi},\ref{eq:truePhi},\ref{eq:trueOmega}),
followed by appropriate re-scaling of the conjugate order
parameters, these three functions acquire the following form
(modulo irrelevant constants):
\begin{eqnarray}
\Psi &=&i\del^2 \!\!\sum_{\ell\ell^\prime\leq t_{\rm max}/\del}
\Big[\hC(\ell,\ell^\prime)C(\ell,\ell^\prime)+\hK(\ell,\ell^\prime)K(\ell,\ell^\prime)+\hL(\ell,\ell^\prime)L(\ell,\ell^\prime)\Big]
~~~~ \label{eq:contPsi}
\\
\Phi&=&\frac{1}{N}\log
 \bigbra\int\!
\D A\D\hA~\W[A,\hat{A}|Z]
 \bigket_{\!\{Z\}}
\label{eq:contPhi}
\\
\Omega&=&\frac{1}{N}\sum_i \log \int\!\prod_{\ell=1}^{t_{\rm max}
/\del}[dq(\ell)d\hq(\ell)]~\bra M_i[\{q,\hq,z\}]\ket_{\bz}
 \label{eq:contOmega}
\end{eqnarray}
with
\be
\W[A,\hat{A}|Z]=e^{i \sum_{\ell=1}^{t_{\rm max}/\del}
\hA(\ell)[A(\ell)-A_e(\ell)]-\frac{1}{4}\sum_{\ell\ell^\prime=1}^{
t_{\rm max}/\del
}\overline{W}[\ell,\ell^\prime;A,Z]M[\ell,\ell^\prime;A,\hA]}
\label{eq:Ameasure}
 \ee
 It is clear that $\Psi$ and $\Omega$ now have proper
$N\to\infty$ limits. The canonical choice of $\del$ is
subsequently determined by the mathematical condition that
$\lim_{N\to\infty}\Phi[C,K,L]\neq 0$, but finite. It follows that
(\ref{eq:truesaddlepoint}) is again dominated by its physical
saddle-point, and we are nearly back in familiar territory.

 \subsection{The saddle point equations}

 In order to eliminate the fields
$\{\psi_i(\ell),\theta_i(\ell)\}$, and thereby simplify our
equations, we next extract the physical meaning of our order
parameters from the generating functional by taking appropriate
derivatives with respect to these fields. This gives
\begin{eqnarray}
C(\ell,\ell^\prime)&=& \lim_{N\to\infty}
\frac{1}{N}\sum_i\overline{\bra
s_i(\ell)s_i(\ell^\prime)\ket}=\bra
\sigma[q(\ell),z(\ell)]\sigma[q(\ell^\prime),z(\ell^\prime)]\ket_\star
\label{eq:identifyC}
\\
G(\ell,\ell^\prime)&=&\lim_{N\to\infty}\frac{1}{N}\sum_i\frac{\partial\overline{\bra
s_i(\ell)\ket}}{\partial\theta_i(\ell^\prime)} =-i\bra
\sigma[q(\ell),z(\ell)]\hq(\ell)\ket_\star
 \label{eq:identifyK}
\\
 0&=&\lim_{N\to\infty}\frac{1}{N}\sum_i\frac{\partial^2
1}{\partial\theta_i(\ell)\partial\theta_i(\ell^\prime)} =-\bra
\hq(\ell)\hq(\ell^\prime)\ket_\star
 \label{eq:identifyL}
\end{eqnarray}
Thus at the physical saddle-point of (\ref{eq:truesaddlepoint}) we
 have the usual relations $L(\ell,\ell^\prime)=0$ and
$K(\ell,\ell^\prime)=iG(\ell,\ell^\prime)$, where $G$ denotes the
single-site response function.
 Upon varying $\{\hC,\hK,\hL\}$ in (\ref{eq:truesaddlepoint}) we
 reproduce
self-consistently the by now standard equations
\begin{eqnarray}
C(\ell,\ell^\prime)&=&\bra
\sigma[q(\ell),z(\ell)]\sigma[q(\ell^\prime),z(\ell^\prime)]\ket_\star
\label{eq:SPmem1}\\[1mm]
 G(\ell,\ell^\prime)&=&-i\bra
\sigma[q(\ell),z(\ell)] \hq(\ell^\prime)\ket_\star
\label{eq:SPmem2}
\\[1mm]
L(\ell,\ell^\prime)&=&\bra \hq(\ell) \hq(\ell^\prime) \ket_\star=0
\label{eq:SPmem3}
\end{eqnarray}
 We  turn to
variation of the order parameters $\{C,K,L\}$ in $\Psi+\Phi$ (as
$\Omega$ only depends on the conjugate order parameters). In
working out derivatives of $\Phi$ we observe that the conjugate
bids effectively act as differential operators, i.e. $\hA(s)\to
i\partial/\partial A_e(s)$.  This gives us our remaining three
saddle point equations:
\begin{eqnarray}
\hspace*{-15mm} \hC(s,s^\prime)&=&
\lim_{N\to\infty}\frac{i}{4N\del^2}~\frac{
\frac{\partial^2}{\partial A_e(s)\partial A_e(s^\prime)}
\bigbra\int\!\D A\D\hA~\W[A,\hat{A}|Z]~
 \overline{W}[s,s^\prime;A,Z]
 \bigket_{\!\{Z\}}}
 {\bigbra
\int\!\D A\D\hA~\W[A,\hat{A}|Z]
 \bigket_{\!\{Z\}}}
 \label{eq:SPhC}
\\
\hspace*{-15mm} \hK(s,s^\prime)&=&\lim_{N\to\infty}
\frac{-\rate}{2N\del^2} ~\frac{ \frac{\partial}{\partial
A_e(s)}\bigbra\int\! \D A\D\hA~\W[A,\hat{A}|Z]~
\overline{W}[s,s^\prime;A,Z] A(s^\prime)
 \bigket_{\!\{Z\}}}
 {\bigbra
\int\! \D A\D\hA~\W[A,\hat{A}|Z]
 \bigket_{\!\{Z\}}}
 \label{eq:SPhK}
 \\
 \hspace*{-15mm}
\hL(s,s^\prime)&=&\lim_{N\to\infty} \frac{-i\rate^2}{4N\del^2}
~\frac{ \bigbra\int\! \D A\D\hA~\W[A,\hat{A}|Z]~
\overline{W}[s,s^\prime;A,Z]A(s)A(s^\prime)
 \bigket_{\!\{Z\}}}
 {\bigbra
\int\!\D A\D\hA~\W[A,\hat{A}|Z]
 \bigket_{\!\{Z\}}}
 \label{eq:SPhL}
\end{eqnarray}
At the physical saddle-point, we may use $L=0$ and the symmetry of
$\overline{W}[\ldots]$ to simplify the function
$M[\ell,\ell^\prime;A,\hat{A}]$ which occurs in the measure
(\ref{eq:Ameasure}) to
\begin{equation}
 M[\ell,\ell^\prime;A,\hA]=
 \hA(\ell)[1+C(\ell,\ell^\prime)]\hA(\ell^\prime)
 -2i\rate\hA(\ell)G(\ell,\ell^\prime)A(\ell^\prime)
 \end{equation}
The generating fields $\{\tilde\psi_i(\ell)\}$ are now no longer
needed and can be removed. The perturbations $\tilde{\theta}_i$
are still useful for calculating the response function $G$, but
can be chosen site-independent, i.e.
$\tilde{\theta}_i(\ell)=\tilde\theta(\ell)$. The measure
(\ref{eq:measureMi}) will then lose its site dependence. Also the
functions $\{\Psi,\Phi,\Omega\}$ have at this stage become
obsolete. We may define a new time $t=\ell\del=\order(N^0)$, which
will be real-valued as $N\to\infty$, and we may take the limit
$N\to\infty$ in the definitions of our observables. The latter can
subsequently be written in terms of the new real-valued time
arguments, $C(\ell,\ell^\prime)\to C(t,t^\prime)$ (and similar for
the other kernels).

\section{The resulting theory}

\subsection{Simplification of saddle-point equations}

 We may now summarize our saddle-point equations for $\{C,G\}$ in the usual
compact way, in terms of an effective single agent process:
\begin{equation}
C(t,t^\prime)=\bra \sgn[q(t)]\sgn[q(t^\prime)]\ket_\star ~~~~~~~~
G(t,t^\prime)=-i\bra \sgn[q(t)]\hq(t^\prime)\ket_\star
\label{eq:SPCG}
\end{equation}
with a measure which is defined in terms of path integrals, as in
\cite{CoolHeim01} (and with time integrals running from $t=0$ to
$t=t_{\rm max}$):
\begin{eqnarray}
 \bra
g[\{q,\hq,z\}]\ket_\star&=& \frac{ \int\{dq d\hq\}~\bra
M[\{q,\hq,z\}]~g[\{q,\hq,z\}]\ket_{\bz}}{ \int\{dqd\hq\}~\bra
M[\{q,\hq,z\}]\ket_{\bz}}
\\[1mm]
M[\{q,\hq,z\}]&=&
p_0(q(0))~e^{i\int\!dt~\hq(t)\left[\frac{d}{dt}q(t)-\theta(t)-\int\!dt^\prime
\hK(t^\prime,t)\sigma[q(t^\prime),z(t^\prime)]\right]} \nonumber\\
&& \hspace*{2mm} \times~e^{-i\int\!dtdt^\prime\left[
\hL(t,t^\prime)\hq(t)\hq(t^\prime)
+\hC(t,t^\prime)\sigma[q(t),z(t)]\sigma[q(t^\prime),z(t^\prime)]\right]}
~~~~~ \label{eq:nearly_final_measure}
\end{eqnarray}
To find the kernels $\{\hC,\hK,\hL\}$ we have to evaluate
equations (\ref{eq:SPhC},\ref{eq:SPhK},\ref{eq:SPhL}) further,
remembering that the left-hand sides as yet still involve the
integer time labels $(s,s^\prime)$, rather  than the  continuous
times. Now the scaling chosen for $\del$ with $N$ which we adopt
will be crucial. We observe that all complications are contained
in the evaluation, for large $N$ and for any given realization of
the fake market information path $\{Z\}$, of objects of the
following general form (with all operators evaluated at the
saddle-point):
\begin{equation}
\bra Q[\{A\}]\ket_{\{A|Z\}}= \int\!\D
A\D\hA~\W[A,\hat{A}|Z]~Q[\{A\}]
 \label{eq:defineAmeasure}
\end{equation}
We can confirm, by repeating the steps taken in evaluating the
disorder-averaged generating functional $\overline{Z[\bpsi]}$ but
now for calculating averages of arbitrary functions of the overall
market bid path $\{A\}$, that the physical interpretation of the
measure (\ref{eq:defineAmeasure}) is
\begin{equation}
\lim_{N\to\infty}\overline{\bra Q[\{A\}]\ket}=\bigbra\room\bra
Q[\{A\}]\ket_{\{A|Z\}}\bigket_{\!\{Z\}} \label{eq:meaningAmeasure}
\end{equation}
Thus (\ref{eq:defineAmeasure}) defines the asymptotic
disorder-averaged probability density for observing a `path'
$\{A\}$ of global bids, for a given realization
 of the fake history  path $\{Z\}$.
To evaluate
 (\ref{eq:defineAmeasure}) we introduce two path-dependent matrices
 $G[A,Z]$ and $D[A,Z]$, with entries
 \begin{eqnarray}
G[A,Z](\ell,\ell^\prime)&=&\overline{W}[\ell,\ell^\prime;A,Z]G(\ell,\ell^\prime)\\
D[A,Z](\ell,\ell^\prime)&=&\overline{W}[\ell,\ell^\prime;A,Z][1+C(\ell,\ell^\prime)]
\end{eqnarray}
Definition (\ref{eq:overlineW}) tells us that
$G[A,Z](\ell,\ell^\prime)=G(\ell,\ell^\prime)$ if the `history'
observed at stage $\ell$ is identical to that observed at stage
$\ell^\prime$, and zero otherwise, and similarly for the relation
between $D[A,Z](\ell,\ell^\prime)$ and $1+C(\ell,\ell^\prime)$. We
now use auxiliary integration variables $\{\phi_\ell\}$ to
linearize the term in the exponent of (\ref{eq:defineAmeasure})
which is quadratic in $\hA$, and use causality of the response
function $G$ where appropriate:
\begin{eqnarray}
\hspace*{-20mm}
 \bra Q[\{A\}]\ket_{\{A|Z\}}&=&  \int\!\prod_{\ell=1}^{t_{\rm max}
/\del}\!\left[\frac{dA(\ell)d\hA(\ell)}{2\pi} e^{i
\hA(\ell)[A(\ell)-A_e(\ell) +
\frac{1}{2}\rate\sum_{\ell^\prime<\ell}
 G[A,Z](\ell,\ell^\prime)A(\ell^\prime)]}
\right]Q[\{A\}] \nonumber \\
 \hspace*{-20mm}
 && ~~~~ \times ~
\frac{\int[\prod_{\ell=1}^{t_{\rm max} /\del}
d\phi_\ell]e^{-\sum_{\ell\ell^\prime=1}^{t_{\rm max}
/\del}\phi_\ell
(D^{-1}[A,Z])_{\ell\ell^\prime}\phi_{\ell^\prime}-i\sum_{\ell=1}^{t_{\rm
max} /\del} \phi_\ell \hA_\ell}} {\int[\prod_\ell
d\phi_\ell]e^{-\sum_{\ell\ell^\prime=1}^{t_{\rm max}
/\del}\phi_\ell
(D^{-1}[A,Z])_{\ell\ell^\prime}\phi_{\ell^\prime}}} \nonumber
 \\
  \hspace*{-20mm}
&=&  \int\!\Big[\prod_{\ell=1}^{t_{\rm max}
/\del}\!\!dA(\ell)\Big]~ Q[\{A\}]\nonumber
\\
 \hspace*{-20mm}
&& \times\bigbra \prod_{\ell=1}^{t_{\rm max} /\del}\!\delta\left[
A(\ell)-A_e(\ell) + \frac{1}{2}\rate\sum_{\ell^\prime<\ell}
 G[A,Z](\ell,\ell^\prime)A(\ell^\prime)-\phi_\ell
 \right]
\bigket_{\!\{\phi|A,Z\}}\nonumber
\end{eqnarray}
Here $\bra \ldots\ket_{\{\phi|A,Z\}}$ refers to averaging over the
zero-average Gaussian fields $\phi_\ell$ with $\{A,Z\}$-dependent
covariance $\bra
\phi_\ell\phi_{\ell^\prime}\ket_{\{\phi|A,Z\}}=\frac{1}{2}D[A,Z](\ell,\ell^\prime)$.
 We conclude from our expression for $\bra Q[\{A\}]\ket_{\{A|Z\}}$
 that the conditional disorder-averaged probability density  $\cP[\{A\}|\{Z\}]$
 for finding a bid path $\{A\}$, given a realization $\{Z\}$ of the pseudo-history,  is given by
\begin{eqnarray}
\hspace*{-25mm} \cP[\{A\}|\{Z\}] &=&
 \bigbra
\prod_{\ell=1}^{t_{\rm max} /\del}\!\delta\left[ A(\ell)-A_e(\ell)
+ \frac{1}{2}\rate\sum_{\ell^\prime<\ell}
 G(\ell,\ell^\prime)\overline{W}[\ell,\ell^\prime;A,Z]A(\ell^\prime)-\phi_\ell
 \right]
\bigket_{\!\!\{\phi|A,Z\}}\nonumber \\ \hspace*{-25mm}&&
\label{eq:Astats}
\end{eqnarray}
with $\bra Q[\{A\}]\ket_{\{A|Z\}}=\int\![\prod_\ell
dA(\ell)]\cP[\{A\}|\{Z\}]Q[\{A\}]$. Causality ensures that the
density (\ref{eq:Astats}) is normalized, since both $\phi_\ell$
and $G[A,Z](\ell,\ell^\prime)$ involve only entries of the paths
$\{A,Z\}$ with times $k<\ell$.

Having established (\ref{eq:Astats}), our equations
(\ref{eq:SPhC},\ref{eq:SPhK},\ref{eq:SPhL}) can be simplified
considerably. We immediately find that $\hC= 0$.
 To simplify comparison with the theory of \cite{CoolHeim01}
 (corresponding to $\zeta=1$), we will make a final change in
notation and put
\begin{equation} \hK(\ell,\ell^\prime)=-\alpha
R(\ell^\prime,\ell)~~~~~~~~\hL(\ell,\ell^\prime)=-\frac{1}{2}\alpha
i\Sigma(\ell,\ell^\prime) \end{equation} This allows us, with
$p=\alpha N$ and in anticipation of our expected time scaling
$\del=\rate/2p$ (known from the analysis in \cite{CoolHeim01} of
the Markovian limit $\zeta=1$), to write the remaining equations
(\ref{eq:SPhK},\ref{eq:SPhL}) in the simple form
\begin{eqnarray}
R(\ell,\ell^\prime) &=& \lim_{N\to\infty} \frac{\partial}{\partial
A_e(\ell^\prime)} \left\{ \frac{\rate}{2p\del^2} \bigbra \bra
\overline{W}[\ell^\prime,\ell;A,Z] A(\ell)\ket_{\{A|Z\}}
 \bigket_{\!\{Z\}}\right\}
 \label{eq:finalR}
 \\
\Sigma(\ell,\ell^\prime) &=& \lim_{N\to\infty}
\left\{\frac{\rate^2}{2p\del^2} \bigbra \bra
\overline{W}[\ell,\ell^\prime;A,Z] A(\ell)A(\ell^\prime)
\ket_{\{A|Z\}} \bigket_{\!\{Z\}} \right\}
 \label{eq:finalSigma}
\end{eqnarray}
We see that $R$  defines a response function associated with
external bid perturbation, and hence obeys causality:
$R(\ell,\ell^\prime)=0$ for $\ell^\prime>\ell$. This, in turn,
enables us to simplify equations (\ref{eq:SPCG}) for $\{C,G\}$ and
the measure $\bra\ldots\ket_\star$
 to a form identical to that found in \cite{CoolHeim01} for the Markovian (`fake history')
 on-line MG:
\begin{eqnarray}
C(t,t^\prime)&=&\bra
\sigma[q(t),z(t)]\sigma[q(t^\prime),z(t^\prime)]\ket_\star
\label{eq:finalmemC}
\\
G(t,t^\prime)&=&\frac{\delta}{\delta\tilde\theta(t^\prime)}\bra
\sigma[q(t),z(t)]\ket_\star \label{eq:finalmemG}
\\
 \bra
g[\{q,z\}]\ket_\star&=& \frac{ \int\!\{dq\}~\bra
g[\{q,z\}]~M[\{q,z\}]\ket_{\bz}}{ \int\!\{dq\}~\bra
M[\{q,z\}]\ket_{\bz}}
\\
M[\{q,z\}]&=& p_0(q(0))~\int\!\{d\hq\}~e^{
-\frac{1}{2}\alpha\int\!dtdt^\prime~
\Sigma(t,t^\prime)\hq(t)\hq(t^\prime)} \nonumber
\\
&& \times~e^{
i\int\!dt~\hq(t)\left[\frac{d}{dt}q(t)-\tilde\theta(t)+\alpha\int\!dt^\prime~
R(t,t^\prime)\sigma[q(t^\prime),z(t^\prime)]\right] }
\label{eq:final_memmeasure}
\end{eqnarray}

\subsection{Summary and interpretation}

 We recognize that (\ref{eq:final_memmeasure}) describes the
usual effective single-trader equation with a retarded
self-interaction and zero-average Gaussian noise $\eta(t)$ with
covariances $\bra \eta(t)\eta(t^\prime)\ket=\Sigma(t,t^\prime)$:
\begin{eqnarray}
\frac{d}{dt}q(t)&=& \tilde{\theta}(t)-\alpha\int_0^t\!dt^\prime~
R(t,t^\prime)~\sigma[q(t^\prime)] +\sqrt{\alpha}~\eta(t)
\label{eq:single_trader_mem}
\end{eqnarray}
We have used the fact, as in \cite{CoolHeim01}, that the
discontinuity of the correlation function for equal times, i.e.
$C(t,t)=1$, will in the continuous time limit be irrelevant. This
implies that we may carry out the averages over the decision noise
and are left only with expressions involving
$\sigma[q]=\int\!dz~P(z)\sigma[q,z]$, and that (with the exclusion
of $t=t^\prime$, where one has $C(t,t)=1$) the order parameter
equations (\ref{eq:finalmemC},\ref{eq:finalmemG}) simplify to
\be
C(t,t^\prime)=\bra \sigma[q(t)]\sigma[q(t^\prime)]\ket_\star
~~~~~~~~
G(t,t^\prime)=\frac{\delta}{\delta\tilde{\theta}(t^\prime)}\bra
\sigma[q(t)]\ket_\star \label{eq:finalmemCG}
 \ee
 Our remaining problem is to solve the order
parameters $\{R,\Sigma\}$ from
(\ref{eq:finalR},\ref{eq:finalSigma}). To do so we must select the
canonical time scale $\del$ such that the $N\to\infty$ limit in
(\ref{eq:finalR},\ref{eq:finalSigma}) is both non-trivial (i.e.
$\del$ sufficiently small) and well-defined (i.e. $\del$ not too
small). For the special value $\zeta=1$ we know \cite{CoolHeim01}
that $\del=\rate/2p$. Although here we have followed a different
route towards a continuous time description, we show in
\ref{app:fake_only} that indeed $\del=\rate/2p$, by working out
our present equations in detail for the fake history limit
$\zeta\to 1$. Given this canonical time scaling and given the
definition
$\overline{W}[\ell,\ell^\prime;A,Z]=\delta_{\blambda(\ell,A,Z),\blambda(\ell^\prime,A,Z)}$,
we find our equations (\ref{eq:finalR},\ref{eq:finalSigma}) taking
their final forms:
\begin{eqnarray}
\room R(t,t^\prime) &=& \lim_{\del\to 0}~\frac{\delta}{\delta
A_e(t^\prime)} ~\left. \bigbra\!\bigbra  A(\ell)
\delta_{\blambda(\ell,A,Z),\blambda(\ell^\prime,A,Z)}
 \bigket\!\bigket_{\!\{A,Z\}}\right|_{\ell=t/\del,\ell^\prime=t^\prime/\del}~~~~
 \label{eq:finallR}
 \\
 \room
\Sigma(t,t^\prime) &=& \rate ~\lim_{\del\to 0} ~\frac{1}{\del}
\left.\bigbra\!\bigbra  A(\ell)A(\ell^\prime)
\delta_{\blambda(\ell,A,Z),\blambda(\ell^\prime,A,Z)} \bigket\!
\bigket_{\!\{A,Z\}}\right|_{\ell=t/\del,\ell^\prime=t^\prime/\del}~~~~
 \label{eq:finallSigma}
\end{eqnarray}
with $\delta/\delta A_e(\ell)=\del^{-1}\!\partial/\partial
A(\ell)$. Here $\bra\bra \ldots\ket\ket_{A,Z}$ refers to an
average over the stochastic process (\ref{eq:Astats}) for the
overall bids $\{A\}$ and over the pseudo-history $\{Z\}$. The bid
evolution process can be written in more explicit form as
\begin{eqnarray}
 A(\ell)&=& A_e(\ell)+\phi_\ell -
\frac{1}{2}\rate\sum_{\ell^\prime<\ell}
 G(\ell,\ell^\prime)\delta_{\blambda(\ell,A,Z),\blambda(\ell^\prime,A,Z)}~A(\ell^\prime)
\label{eq:effective_bid_process}
\end{eqnarray}
with the zero-average Gaussian random fields $\{\phi\}$,
characterized by \begin{eqnarray} \bra
\phi_\ell\phi_{\ell^\prime}\ket_{\{\phi|A,Z\}}&=&\frac{1}{2}[1+C(\ell,\ell^\prime)]
~\delta_{\blambda(\ell,A,Z),\blambda(\ell^\prime,A,Z)}
\label{eq:finall_phistats}
\end{eqnarray}
Equation (\ref{eq:effective_bid_process}) is to be interpreted as
follows. For every realization $\{Z\}$ of the fake history `path'
one iterates (\ref{eq:effective_bid_process}) to find  successive
bid values upon generating the zero-average Gaussian random
variables $\phi_\ell$ with statistics (\ref{eq:finall_phistats})
(which depend, in turn, on the recent bid realizations). The
result is averaged over the fake history paths $\{Z\}$.

Let us now summarize the structure of the present theory
describing the MG with true market history in the limit
$N\to\infty$, by indicating the similarities and the differences
with the previous theory describing the on-line MG without market
history:\vspace*{1mm}
\begin{description}
\item[{\sl similarities between the theory of real and fake history MGs:}]$~$\\[-4mm]
\begin{itemize}
\item
The MG with real history is described again by the effective
single agent equation (\ref{eq:single_trader_mem}), from which the
usual order dynamical order parameters $\{C,G\}$ are to be solved
self-consistently via (\ref{eq:finalmemCG}).
\item
The scaling with $N$ of the characteristic times in the MG with
history is identical to that of the MG without history, if we
avoid highly biased global bid initializations (where the MG with
history acts faster by a factor $\sqrt{N}$).
\end{itemize}
\item[{\sl differences between the theory of real and fake history MGs:}]$~$\\[-4mm]
\begin{itemize}
\item
Real and fake history MGs differ in the retarded self-interaction
kernel $R$ and the noise covariance kernel $\Sigma$ of the single
agent equation. Without history, $\{R,\Sigma\}$ were found as
explicit functions of $\{C,G\}$. With history they are to be
solved from an effective equation (\ref{eq:effective_bid_process})
for the evolving global bid.
\end{itemize}
\clearpage
\item[{\sl the effective global bid process:}]$~$\\[-4mm]
\begin{itemize}
\item
The effective global bid process (\ref{eq:effective_bid_process})
is itself independent of the stochastic effective single trader
process (\ref{eq:single_trader_mem}). The two are linked only via
the (time dependent) order parameters occurring in their
definitions.
\item
At each stage in the process (\ref{eq:effective_bid_process}), the
bid $A(\ell)$ is coupled directly only to bids in the past at
times $\ell^\prime$ with {\em identical realization of the $M$-bit
history string}. In addition, only those effective global bid
noise variables $\phi_\ell$ are correlated which correspond to
times $\ell$ with identical realizations of the $M$-bit history
string.
\end{itemize}
\end{description}
 The differences
between the two `fake history' definitions
(\ref{eq:typeA},\ref{eq:typeB}) (i.e. consistent versus
inconsistent) are seen to be limited to the details of the
averaging process $\bra \ldots\ket_{\{Z\}}$.

In \ref{app:fake_only} we show how one can recover from
(\ref{eq:single_trader_mem},\ref{eq:effective_bid_process}) the
earlier theory of \cite{CoolHeim01} in the fake history limit
$\zeta\to 1$. This exercise serves two purposes. Firstly, it
confirms that the canonical time scale of our process is indeed
given by $\del=\rate/2p$ (modulo an irrelevant multiplicative
constant). More importantly, being the simplest instance of our
presently studied class of MG models, it provides useful intuition
on how we might proceed to find solutions of our effective
processes
(\ref{eq:single_trader_mem},\ref{eq:effective_bid_process}) in the
general case.

\section{The role of history statistics}

We continue with our analysis of the full MG with history, and
next show that all the effects induced by having real market
history can be concentrated in the statistics of the $M$-bit
memory strings $\blambda$ of (\ref{eq:mem_lambda}). More
specifically, the core objects in the theory will turn out to be
the following functions, which measure the joint probability to
find identical histories in the effective global bid process
(\ref{eq:effective_bid_process}) at $k$ specified times
$\{\ell_1,\ldots,\ell_k\}$, relative to the probability $p^{-k}$
for this to happen in the case of randomly drawn fake histories
and non-identical times:
\begin{eqnarray}
\Delta_k(\ell_1,\ldots,\ell_k)&=& p^{k-1} \sum_{\blambda}
\big\bra\!\big\bra ~\prod_{i=1}^k
\delta_{\blambda,\blambda(\ell_i,A,Z)}~\big\ket\!\big\ket_{\!\{A,Z\}}
\label{eq:history_stats}
\end{eqnarray}
We have abbreviated
$\sum_{\blambda}=\sum_{\blambda\in\{-1,1\}^M}$, with $2^M=p=\alpha
N$. For any value of $k$ one recovers in the random history limit
and for non-identical times $\lim_{\zeta\to 1}\Delta_k(\ldots)=1$.
For $k=1$ one has $\Delta_1(\ell)= \sum_{\blambda} \bra\!\bra
\delta_{\blambda,\blambda(\ell,A,Z)}\ket\!\ket_{\{A,Z\}}=1$, for
any $\zeta$. In contrast, for arbitrary $\zeta$ (i.e. when
allowing for real histories) and $k>1$ the functions
(\ref{eq:history_stats}) are nontrivial.

\subsection{Reduction of the kernels $\{R,\Sigma\}$}

We will follow as much as possible the steps which we took in
\ref{app:fake_only} in order to recover the $\zeta=1$ equations
(\ref{eq:nomemRnew},\ref{eq:nomemSigmanew}). We re-write the
global bid equation (\ref{eq:effective_bid_process}) as
\begin{eqnarray*}
\sum_{\ell^\prime\leq \ell}\Big\{ \delta_{\ell\ell^\prime} +
\frac{1}{2}\rate
 G(\ell,\ell^\prime)\delta_{\blambda(\ell,A,Z),\blambda(\ell^\prime,A,Z)}
 \Big\}A(\ell^\prime)
&=& A_e(\ell)+\phi_\ell
\end{eqnarray*}
and we formally invert the operator on the left-hand side, using
$\del=\rate/2p$:
\begin{eqnarray}
A(\ell)&=&
A_e(\ell)+\phi_\ell+\sum_{r>0}(-\frac{\rate}{2})^r\sum_{\ell_1\ldots
\ell_r} G(\ell,\ell_1)G(\ell_1,\ell_2)\ldots G(\ell_{r-1},\ell_r)
\nonumber \\ &&\times \Big[\prod_{i=1}^r
\delta_{\blambda(\ell,A,Z),\blambda(\ell_i,A,Z)} \Big] ~
\big[A_e(\ell_r)+\phi_{\ell_r}\big] \label{eq:Aexplicit}
\end{eqnarray}
Expression (\ref{eq:Aexplicit}) is itself not yet a solution of
(\ref{eq:effective_bid_process}), since the bids $\{A(s)\}$ also
occur inside the history strings $\blambda(\ell^\prime,A,Z)$ at
the right-hand side. We now insert (\ref{eq:Aexplicit}) first into
(\ref{eq:finallR}), and consider only infinitesimal external bid
perturbations $A_e$, so that we need not worry about indirect
effects on $A(\ell)$ of these perturbations via the history
strings $\blambda(s,A,Z)$:
\begin{eqnarray}
\hspace*{-10mm} \room R(t,t^\prime) &=&
\delta(t-t^\prime)+\lim_{\del\to 0} \left\{\sum_{r>0}(-\del)^{r-1}
\sum_{\ell_1\ldots \ell_{r-1}}\!
G(\ell,\ell_1)G(\ell_1,\ell_2)\ldots G(\ell_{r-1},\ell^\prime)
\right.\nonumber
\\ \hspace*{-10mm}&&
\left. \times~ p^r \Big\bra\!\Big\bra
\delta_{\blambda(\ell,A,Z),\blambda(\ell^\prime,A,Z)}\prod_{i=1}^{r-1}
\delta_{\blambda(\ell,A,Z),\blambda(\ell_i,A,Z)}
 \Big\ket\!\Big\ket_{\!\{A,Z\}}\right\}
 \Big|_{\ell=\frac{t}{\del},\ell^\prime=\frac{t^\prime}{\del}}
 \nonumber
 \\
 \hspace*{-10mm}
 &=&
\delta(t-t^\prime)+\lim_{\del\to 0}
\left\{\sum_{r>0}(-\del)^{r-1}\! \sum_{\ell_1\ldots \ell_{r-1}}
G(\ell_0,\ell_1)\ldots G(\ell_{r-1},\ell_r) \nonumber \right.
\\
\hspace*{-10mm} && \left. \hspace*{40mm}\times~
\Delta_{r+1}(\ell_0,\ldots,\ell_{r}) \room\right\}
\Big|_{\ell_0=\frac{t}{\del},\ell_r=\frac{t^\prime}{\del}}
 \label{eq:workoutR}
\end{eqnarray}
Similarly we can insert
 (\ref{eq:Aexplicit})  into
(\ref{eq:finallSigma}), again with $A_e\to 0$, and find
\begin{eqnarray}
\Sigma(t,t^\prime) &=& \rate \lim_{\del\to 0} \frac{1}{\del}
\left\{ \sum_{r,r^\prime\geq 0}(-\del)^{r+r^\prime}
\sum_{\ell_1\ldots \ell_r}G(\ell_0,\ell_1)\ldots
G(\ell_{r-1},\ell_r) \nonumber\right.
\\
&&\left. \times~\sum_{\ell^\prime_1\ldots \ell^\prime_r}
G(\ell^\prime_0,\ell^\prime_1)\ldots
G(\ell^\prime_{r-1},\ell^\prime_r)~p^{r+r^\prime}
 \Big\bra\!\Big\bra~
 \bra\phi_{\ell_r}
 \phi_{\ell^\prime_{r^\prime}}\ket_{\{\phi|A,Z\}}
 \nonumber \right.
\\ &&
\left.\hspace*{-10mm} \times~ \Big[\prod_{i=1}^r
\delta_{\blambda(\ell_0,A,Z),\blambda(\ell_i,A,Z)} \Big]
\Big[\prod_{j=1}^{r^\prime}
\delta_{\blambda(\ell_0^\prime,A,Z),\blambda(\ell^\prime_j,A,Z)}
\Big]~
 \Big\ket\!
\Big\ket_{\!\{A,Z\}}\right\}\Big|_{\ell_0=\frac{t}{\del},\ell^\prime_0=\frac{t^\prime}{\del}}
\nonumber
\\
&=&
 \lim_{\del\to 0} \left\{ \sum_{r,r^\prime\geq
0}(-\del)^{r+r^\prime} \sum_{\ell_1\ldots
\ell_r}G(\ell_0,\ell_1)\ldots G(\ell_{r-1},\ell_r)
\nonumber\right.
\\
&&\left.\hspace*{20mm} \times~\sum_{\ell^\prime_1\ldots
\ell^\prime_r} G(\ell^\prime_0,\ell^\prime_1)\ldots
G(\ell^\prime_{r-1},\ell^\prime_r)~\big[1+C(\ell_r,\ell^\prime_{r^\prime})\big]
 \nonumber \right.
\\ &&
\left.\hspace*{20mm} \times~
\Delta_{r+r^\prime+2}(\ell_0,\ldots,\ell_r,\ell^\prime_0,\ldots,\ell^\prime_{r^{\prime}})
\room\right\}\Big|_{\ell_0=\frac{t}{\del},\ell^\prime_0=\frac{t^\prime}{\del}}
 \label{eq:workoutSigma}
\end{eqnarray}
The limits $\del\to 0$ in
(\ref{eq:workoutR},\ref{eq:workoutSigma}) are well-defined, since
each time summation combines with a factor $\del$ to generate an
integral, whereas pairwise identical times in
(\ref{eq:workoutSigma}) leave a `bare' factor $\del$ but will also
cause $\Delta_{r+r^\prime+2}(\ldots)$ to gain a factor
$p=\rate/2\del$ in compensation.

Since the single agent process (\ref{eq:single_trader_mem}) is
linked to the global bid process (\ref{eq:effective_bid_process})
only via the kernels $\{R,\Sigma\}$, we conclude from
(\ref{eq:workoutR},\ref{eq:workoutSigma}) (which are still fully
exact) that the effects of having true market history are
concentrated solely in the resulting history statistics as
described by the functions (\ref{eq:history_stats}). More
specifically, there is no need for us to solve the global bid
process
 (\ref{eq:effective_bid_process})
beyond knowing the history statistics which it generates.

\subsection{Time-translation invariant stationary states}

In fully ergodic and time-translation invariant  states without
anomalous response, we could in `fake history' MG versions find
exact closed equations for persistent order parameters without
having to solve for the kernels $\{C,G\}$ in full, and locate
phase transitions exactly. This suggests that the same may be true
for MGs with true history. Thus we make
 the standard time-translation invariance (TTI) ansatz for the kernels
 in
 (\ref{eq:single_trader_mem}) and for the correlation- and response functions:
\begin{eqnarray*}
C(t,t^\prime)&=&C(t-t^\prime)~~~~~~~~G(t,t^\prime)=G(t-t^\prime)
\\
R(t,t^\prime)&=&R(t-t^\prime)~~~~~~~~\Sigma(t,t^\prime)=\Sigma(t-t^\prime)
\end{eqnarray*}
 with
 $\chi=\int_0^\infty\!dt~R(t)$ finite.
 It turns out that several
relations between persistent observables in TTI stationary states
of the present non-Markovian MG process, if such states again
exist, can be established on the basis of
  (\ref{eq:single_trader_mem}) alone.
 Upon following established notation conventions and abbreviating time averages as
 $\overline{f}=\lim_{\tau\to\infty}\tau^{-1}\int_0^\tau\!dt~f(t)$,
 we may write the time average of (\ref{eq:single_trader_mem})
as
\begin{eqnarray}
\overline{dq/dt}&=& \overline{\tilde{\theta}}-\alpha \chiR~
\overline{\sigma} +\sqrt{\alpha}~\overline{\eta}
\label{eq:single_trader_timeaverage}
\end{eqnarray}
with $\chiR=\int_0^\infty\!dt ~R(t)$. We may now define the
familiar  effective agent trajectories corresponding to fickle
versus frozen agents as those with either $\overline{dq/dt}=0$ or
$\overline{dq/dt}\neq 0$, respectively. For frozen agents,
consistency demands that
$\sgn[\overline{\sigma}]=\sgn[\overline{dq/dt}]$. It then follows
from (\ref{eq:single_trader_timeaverage}) that the (at least for
$\chiR>0$ complementary and mutually exclusive) conditions for
having a `fickle' or a `frozen' solution can be written as
follows:
\begin{eqnarray}
{\rm fickle:}&~~~
|\overline{\theta}+\sqrt{\alpha}\overline{\eta}|\leq \alpha
\chiR\sigma[\infty],&~~~
\overline{\sigma}=\frac{\overline{\tilde{\theta}}+\sqrt{\alpha}~\overline{\eta}}{\alpha\chiR}
 \label{eq:condition_fickle}
\\
{\rm frozen:}&~~~
|\overline{\theta}+\sqrt{\alpha}\overline{\eta}|>
\alpha\chiR\sigma[\infty] ,&~~~
\overline{\sigma}=\sigma[\infty].\sgn\Big[\frac{\overline{\tilde{\theta}}+\sqrt{\alpha}~\overline{\eta}}{\alpha\chiR}\Big]
\label{eq:condition_frozen}
\end{eqnarray}
Which solution of (\ref{eq:condition_fickle}) and
(\ref{eq:condition_frozen}) we will find depends on the
realization of the noise term $\overline{\eta}$, which is a frozen
Gaussian variable with zero expectation value and with variance
\begin{eqnarray}
S_0^2&=& \bra \overline{\eta}^2\ket_\star=\lim_{\tau\to
\infty}\frac{1}{\tau^2}\int_0^\tau\!dt dt^\prime~
\Sigma(t,t^\prime)=\Sigma(\infty)
\end{eqnarray}
We may now proceed as in \ref{app:fake_only} towards the
calculation of the persistent order parameters $\phi$, $\chi$ and
$c$, where $\phi$ denotes the fraction of frozen agents in the
stationary state, where $\chi=\int_0^\infty\!dt~G(t)$, and with
\begin{eqnarray}
c&=&
\lim_{t\to\infty}C(t)=\lim_{\tau\to\infty}\frac{1}{\tau^2}\int_0^\tau\!dtdt^\prime~\bra
\sigma[q(t)]\sigma[q(t^\prime)]\ket_\star=\bra
\overline{\sigma}^2\ket_\star
\end{eqnarray}
Upon introducing the short-hand
$u=\sqrt{\alpha}\chiR\sigma[\infty]/S_0\sqrt{2}$, and upon using
the conditions and relations
(\ref{eq:condition_fickle},\ref{eq:condition_frozen}), we find in
the limit  $\tilde{\theta}\to 0$ of vanishing external fields:
\begin{eqnarray}
\phi&=&
\int\!\frac{d\overline{\eta}}{S_0\sqrt{2\pi}}~e^{-\frac{1}{2}\overline{\eta}^2/S_0^2}
\theta\Big[|\overline{\eta}|-\sqrt{\alpha}\chiR\sigma[\infty]\Big]
\nonumber
\\ &=& 1- {\rm Erf}[u] \label{eq:phi_eqn}
\\
c&=&\int\!\frac{d\overline{\eta}}{S_0\sqrt{2\pi}}~e^{-\frac{1}{2}\overline{\eta}^2/S_0^2}\left\{
\room
\theta\Big[|\overline{\eta}|-\sqrt{\alpha}\chiR\sigma[\infty]\Big]
\sigma^2[\infty] \right. \nonumber \\ && \left. \hspace*{40mm} +~
\theta\Big[\sqrt{\alpha}\chiR\sigma[\infty]-|\overline{\eta}|\Big]
\frac{\overline{\eta}^2}{\alpha\chiR^2}\right\} \nonumber
\\
&=&\sigma^2[\infty]\Big\{ 1-{\rm Erf}[u]+\frac{1}{2u^2}{\rm
Erf}[u]-\frac{1}{u\sqrt{\pi}}e^{-u^2}\Big\} \label{eq:c_eqn}
\\
\chi&=&
\int\!\frac{d\overline{\eta}}{S_0\sqrt{2\pi}}~e^{-\frac{1}{2}\overline{\eta}^2/S_0^2}\frac{\partial
\overline{\sigma}}{\partial(\sqrt{\alpha}\overline{\eta})}
\nonumber
\\
&=& {\rm Erf}[u]/\alpha\chiR
 \label{eq:chi_eqn}
\end{eqnarray}
Hence, in order to find the TTI stationary solution
$\{\phi,c,\chi\}$ and the phase transition point (defined by
$\chi\to\infty$), we only need to extract expressions for $\chiR$
and $S_0$ from the stochastic overall bid process
(\ref{eq:effective_bid_process}). Using
 (\ref{eq:workoutR},\ref{eq:workoutSigma}), the latter can be written as
\begin{eqnarray}
\chiR&=& \int_0^\infty\!dt~ R(t)\nonumber\\
 &=&
1+\lim_{\del\to 0} \Big\{ \sum_{r>0}(-\del)^{r}\!
\sum_{\ell_1\ldots \ell_{r}}
G(\ell_1-\ell_2)G(\ell_2-\ell_3)\ldots
G(\ell_{r-1}-\ell_{r})G(\ell_{r}) \nonumber
\\
&&  \hspace*{40mm}\times~ \Delta_{r+1}(\ell_1,\ldots,\ell_{r},0)
\Big\} \label{eq:ChiR}
\\
S_0^2
&=&\lim_{L\to\infty}\frac{1}{L^2}\sum_{\ell_0,\ell^\prime_0\leq L}
 \lim_{\del\to 0} \left\{ \sum_{r,r^\prime\geq
0}(-\del)^{r+r^\prime} \sum_{\ell_1\ldots
\ell_r}G(\ell_0-\ell_1)\ldots G(\ell_{r-1}-\ell_r)
\nonumber\right.
\\
&&\left.\hspace*{20mm} \times~\sum_{\ell^\prime_1\ldots
\ell^\prime_r} G(\ell^\prime_0-\ell^\prime_1)\ldots
G(\ell^\prime_{r-1}-\ell^\prime_r)~\big[1+C(\ell_r-\ell^\prime_{r^\prime})\big]
 \nonumber \right.
\\ &&
\left.\hspace*{20mm} \times~
\Delta_{r+r^\prime+2}(\ell_0,\ldots,\ell_r,\ell^\prime_0,\ldots,\ell^\prime_{r^{\prime}})
\room\right\} \label{eq:Szero}
\end{eqnarray}

\subsection{TTI states with short history correlation times}

Calculating the history statistics kernels
(\ref{eq:history_stats}) from the global bid process
(\ref{eq:effective_bid_process}) is hard, but in those cases where
the history correlation time $L_h$ (measured in individual
iterations $\ell$) in the process is much smaller than $N$, we can
make progress in our analysis of TTI stationary states.
 \begin{figure}[t]
\vspace*{1mm} \hspace*{20mm} \setlength{\unitlength}{0.91mm}
\begin{picture}(90,65)
\put(2,5){\epsfysize=60\unitlength\epsfbox{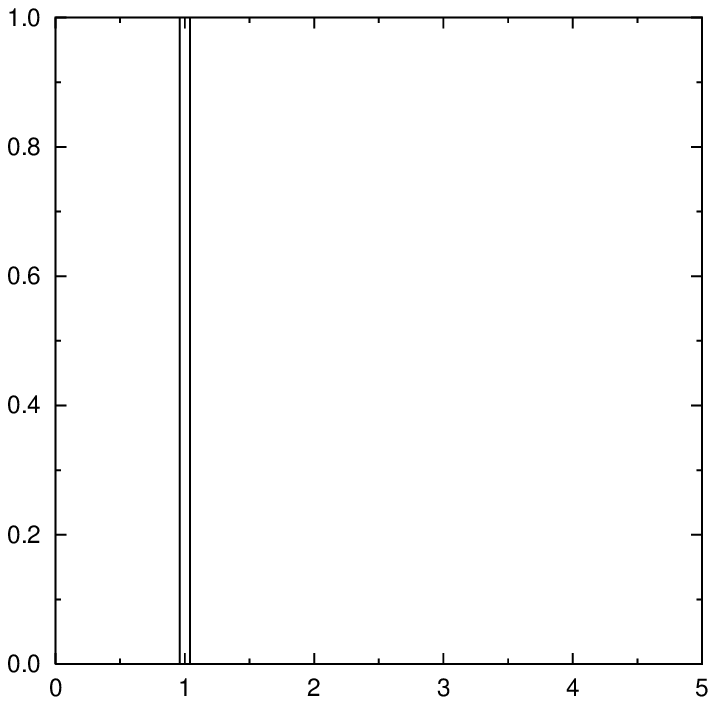}}
\put(37,0){\here{$\freq$}}  \put(-3,35){\here{$\varrho(\freq)$}}

 \put(82,5){\epsfysize=60\unitlength\epsfbox{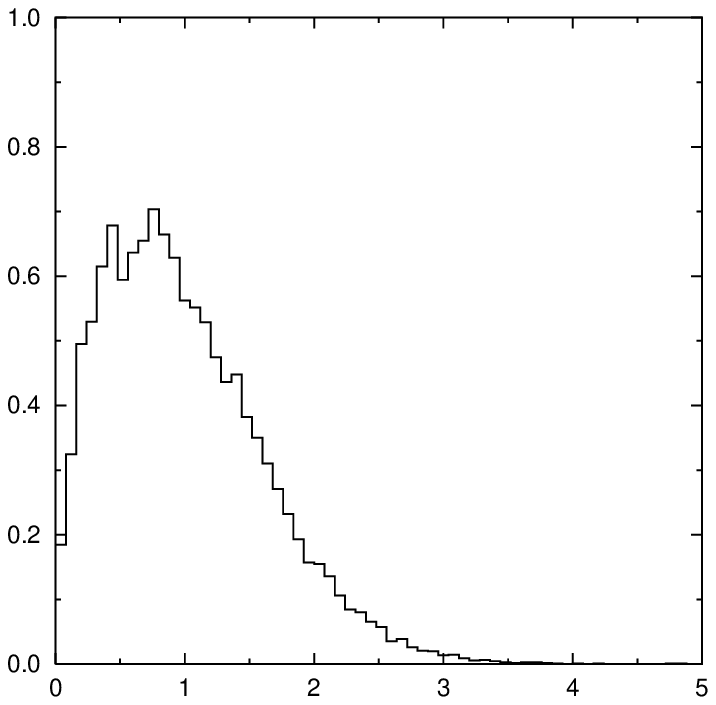}}
\put(117,0){\here{$\freq$}}  \put(77,35){\here{$\varrho(\freq)$}}
\end{picture}
\vspace*{3mm} \caption{Typical examples of history frequency
distributions (\ref{eq:freq_dist})  as measured in simulations of
the on-line MG without decision noise but with full history (i.e.
$\zeta=0$), after equilibration. Here $N=8193$. Left:
$\alpha=0.125$ (in the non-ergodic regime of the MG,  below
$\alpha_c$).  Right: $\alpha=2.0$ (in the ergodic regime,  above
$\alpha_c$). } \label{fig:hist_frequencies}
\end{figure}
We define the asymptotic frequency $\pi_\blambda(A,Z)$ at which
history string $\blambda$ occurs in a given realization $\{A,Z\}$
of our process (\ref{eq:effective_bid_process}) as
\be
\pi_\blambda(A,Z) =\lim_{L\to \infty} \frac{1}{L}\sum_{\ell=1}^L
\delta_{\blambda,\blambda(\ell,A,Z)} \label{eq:hist_freq}
 \ee
Obviously $\sum_{\blambda}\pi_{\blambda}(A,Z)=1$.  For $\zeta=1$
(no history) we would have $\pi_{\blambda}=p^{-1}$ for all
$\blambda$. We may also define the distribution $\varrho(\freq)$
of these asymptotic history frequencies $\pi_{\blambda}(A,Z)$,
relative to the benchmark `no-memory' values $p^{-1}$, and
averaged over the global bid process
(\ref{eq:effective_bid_process}) in the infinite system size (i.e.
continuous time) limit:
\be
\varrho(\freq)=\lim_{p\to\infty} \frac{1}{p}\sum_{\blambda}
\big\bra \!\big\bra \delta\big[\freq-p\pi_{\blambda}(A,Z)\big]
\big\ket\!\big\ket_{\{A,Z\}} \label{eq:freq_dist} \ee
 Our definitions guarantee that
 $\int_0^\infty\!d\freq~\freq \varrho(\freq)=1$ for any $\zeta$.
 For $\zeta=1$ we simply recover $\varrho(\freq)=\delta[\freq-1]$,
 i.e. all histories occur equally frequently.
We have not yet shown that the limit in (\ref{eq:freq_dist})
exists, i.e. that the history frequencies do indeed generally
scale as $\pi_{\blambda}(A,Z)=\order(N^{-1})$. Numerical
simulations, however, confirm quite convincingly that this ansatz
is indeed correct (see e.g. Figure \ref{fig:hist_frequencies}).

 If $L_h$ is the
history correlation time in the process
(\ref{eq:effective_bid_process}), then {\em finite} samples of
history occurrence frequencies can be expected to approach the
asymptotic value (\ref{eq:hist_freq}) as
\be
\frac{1}{2L}\sum_{\ell^\prime=\ell-L}^{\ell+L}\delta_{\blambda,\blambda(\ell^\prime,A,Z)}
=\pi_{\blambda}(A,Z)~\Big[1+\order((L_h/L)^{\frac{1}{2}})\Big]
\label{eq:history_corr_time}
 \ee
  This
implies that in expressions such as (\ref{eq:ChiR}), where
$G(\ell)-G(\ell^\prime)=\order(|\ell-\ell^\prime|/N)$ and where
only time strings $\{\ell_1,\ldots,\ell_k\}$ with mutual temporal
separations of order $\order(N)$ will survive the limit $\del\to
0$, we may choose e.g. $L=\sqrt{L_h N}$ and effectively replace
\be
\Delta_{r+1}(\ell_1,\ldots,\ell_{r},0)~\to~ p^r\sum_{\blambda}
\left[\pi_{\blambda}\big[1+\order(\sqrt{L_h/N})\big]\right]^{r+1}
\ee
 This results in
\begin{eqnarray}
\chiR&=& \lim_{p\to \infty} \frac{1}{p}\sum_{\blambda} \sum_{r\geq
0}(-\chi)^{r}\!\left[p\pi_{\blambda}\big[1+\order(\sqrt{L_h/L})\big]\right]^{r+1}
\nonumber
\\
&=& \lim_{p\to \infty} \frac{1}{p}\sum_{\blambda}
\frac{p\pi_{\blambda}\big[1+\order(\sqrt{L_h/N})\big]}{1+ \chi~
p\pi_{\blambda}\big[1+\order(\sqrt{L_h/N})\big]} =
\int_0^\infty\!d\freq~\varrho(\freq) ~\frac{\freq}{1+ \chi\freq}
\label{eq:simpleChiR}
\end{eqnarray}
(provided indeed $\lim_{N\to\infty}L_h/N=0$). The same
simplification to an expression involving only the distribution
$\varrho(\freq)$ can be achieved in (\ref{eq:Szero}), but there we
have to be more careful in dealing with the occurrences of similar
or identical times in the argument  of (\ref{eq:history_stats}).
We first rewrite (\ref{eq:Szero}) by transforming the iteration
times according to \bd {\rm
for~all}~~i\in\{0,\ldots,r\}:~~~~~~\ell_i=\sum_{j=i}^r s_j \ed
This gives, using $\lim_{s\to\infty}G(s)=0$ (i.e. restricting
ourselves to ergodic states with normal response):
\begin{eqnarray}
\hspace*{-10mm} S_0^2 &=&
 \lim_{\del\to 0}  \sum_{r,r^\prime\geq
0}(-\del)^{r+r^\prime}\! \!\!\sum_{s_0\ldots s_{r-1}>
0}G(s_0)\ldots G(s_{r-1})\!\sum_{s^\prime_0\ldots
s^\prime_{r^\prime\!-1}>0} G(s^\prime_0)\ldots G(s^\prime_{r-1})
\nonumber
\\
\hspace*{-10mm} &&\hspace*{20mm} \times~
\lim_{L\to\infty}\frac{1}{L^2}\sum_{s_r=0}^{L-\sum_{i=0}^{r-1}s_i}\sum_{s^{\prime}_{r^\prime}=0}^{L-\sum_{i=0}^{r^{\prime}-1}s^\prime_i}
\big[1+C(s_r-s^\prime_{r^\prime})\big]
 \nonumber
\\
\hspace*{-10mm} && \hspace*{0mm} \times~
\Delta_{r+r^\prime+2}(s_0+\ldots +s_r
,\ldots,\ell_{r-1}\!+\ell_r,\ell_r,
\ell^\prime_0\!+\ldots+\ell_{r^\prime},\ldots,\ell^\prime_{r^\prime-1}\!+\ell^\prime_{r^\prime},\ell^\prime_{r^\prime})
\nonumber
\end{eqnarray}
Each time summation is compensated either by a factor $\delta_N$
(giving an integral), or limited in range by $L$ and compensated
by an associated factor $L^{-1}$, so that any `pairing' where two
(or more) times are close to each other (relative to the
correlation time $L_h$) will not survive the combined limits
$\del\to 0$ and $L\to\infty$. Thus we may again put
\be
\Delta_{r+r^\prime+2}(\ldots\ldots)~\to~
p^{r+r^\prime+1}\sum_{\blambda}
\left[\pi_{\blambda}\big[1+\order(\sqrt{L_h/N})\big]\right]^{r+r^\prime+2}
\ee
 and find, with $C(\infty)=c$:
\begin{eqnarray}
S_0^2 &=& (1+c)
 \lim_{p\to\infty}  \frac{1}{p}\sum_{\blambda}\sum_{r,r^\prime\geq
0}(-\chi)^{r+r^\prime}
\left[p\pi_{\blambda}\big[1+\order(\sqrt{L_h/N})\big]\right]^{r+r^\prime+2}
\nonumber
\\
&=& (1+c)
 \lim_{p\to\infty}  \frac{1}{p}\sum_{\blambda}
 \frac{\Big[p\pi_{\blambda}\big[1+\order(\sqrt{L_h/N})\big]\Big]^2}{\Big[1+\chi~p\pi_{\blambda}\big[1+\order(\sqrt{L_h/N})\big]\Big]^2}
 \nonumber
\\
&=& (1+c)\int_0^\infty\!d\freq~\varrho(\freq)~
 \frac{\freq^2}{(1+\chi \freq)^2}
 \label{eq:simpleS0}
\end{eqnarray}
Since only $\chiR$ and $S_0$ are needed to solve our effective
single agent process in TTI stationary states, we see that upon
making the ansatz of short history correlation times $L_h\ll N$
 the
effects of history on the persistent order parameters in the MG
are fully concentrated in the distribution $\varrho(\freq)$ of
history frequencies, as defined by  (\ref{eq:freq_dist}). Once
$\varrho(\freq)$ has been extracted from the process
 (\ref{eq:effective_bid_process}), the TTI order
 parameters are given by the solution of the following
 set of
 equations:
\begin{eqnarray}
u&=&\frac{\sigma[\infty]\sqrt{\alpha}\chiR}{S_0\sqrt{2}}~~~~~~~~\chi=
\frac{1-\phi}{\alpha \chiR}~~~~~~~~\phi =1- {\rm Erf}[u]
\label{eq:mem_closed1}
\\
c &=& \sigma^2[\infty]\Big\{1-{\rm Erf}[u]+\frac{1}{2u^2}{\rm
Erf}[u]-\frac{1}{u\sqrt{\pi}}e^{-u^2}\Big\} \label{eq:mem_closed2}
\\
\chiR&=&\int_0^\infty\!d\freq~\varrho(\freq) ~\frac{\freq}{1+
\chi\freq} \label{eq:mem_closed3}
\\
S_0^2&=& (1+c)\int_0^\infty\!d\freq~\varrho(\freq)~
 \frac{\freq^2}{(1+\chi \freq)^2}
 \label{eq:mem_closed4}
\end{eqnarray}
For $\zeta=1$ (the fake history limit) we have
$\varrho(\freq)=\delta[\freq-1]$, leading to $\chiR=(1+\chi)^{-1}$
and $S_0=\sqrt{1+c}/(1+\chi)$, and the above equations are seen to
reduce to the corresponding ones in \cite{CoolHeim01}, as they
should.

\section{Calculating the history statistics}

Upon making the ansatz of short history correlation times in the
MG, we have shown that finding closed equations for persistent TTI
order parameters boils down to calculating the distribution
$\varrho(f)$ of relative history frequencies, as defined in
(\ref{eq:freq_dist})\footnote{A similar conclusion was reached
also in \cite{ChalletMarsili00}, but on the basis of several
approximations. Furthermore, in contrast to the present GFA
approach, in \cite{ChalletMarsili00} there was no way to calculate
$\varrho(f)$ from the theory.}. Our remaining programme of
analysis is: (i) finding an expression for $\varrho(f)$, (ii)
expressing this distribution in terms of the persistent order
parameters $\{c,\phi,\chi,\chiR,S_0\}$, and (iii) confirming
retrospectively the consistency  of assuming short history
correlation times.

\subsection{The moments of $\varrho(f)$}

The distribution  (\ref{eq:freq_dist}) is generated by the
non-Markovian process
 (\ref{eq:effective_bid_process}), which we cannot hope to solve
 directly. However, we can get away with a
 self-consistent calculation which does
 not require solving (\ref{eq:effective_bid_process}) in full.
We focus on the moments $\mu_k$ of the
 distribution $\varrho$, from which the latter can always be
 recovered (if the integrals below exist):
 \begin{eqnarray}
 \mu_k&=& \int_0^\infty\!df~\varrho(f)f^k
 \label{eq:mu_k}
 \\
\varrho(f)&=& \int\!\frac{d\omega}{2\pi}e^{i\omega f} \sum_{k\geq
0}\frac{\mu_k}{k!} (-i\omega)^k
 \label{eq:rho_f}
\end{eqnarray}
Obviously $\mu_0=\mu_1=1$, for any $\zeta$, which follows directly
from definition (\ref{eq:freq_dist}). In the absence of history
(i.e. $\zeta=1$) we have $\varrho(f)=\delta[f-1]$, so that
$\mu_k=1$ for all $k\geq 0$. We will rely on the sum over moments
in (\ref{eq:rho_f}) converging on scales of $k$ which are
independent of $N$. This is equivalent to saying that the limit
(\ref{eq:freq_dist}) is well-defined, so it does not restrict us
further. By combining the definitions
(\ref{eq:hist_freq},\ref{eq:freq_dist},\ref{eq:mu_k}) and
(\ref{eq:mem_lambda}), we can obtain a more explicit but still
relatively simple expression for the moments $\mu_k$:
\begin{eqnarray}
\mu_k &=& \frac{1}{p}\sum_{\blambda}\big\bra \!\big\bra
\big[p\pi_{\blambda}(A,Z)\big]^k
\big\ket\!\big\ket_{\{A,Z\}}\nonumber
\\
&=& \lim_{L\to\infty}\frac{p^k}{L^k}\sum_{\ell_1 \ldots\ell_k=1}^L
\Big\bra \!\Big\bra \prod_{i=1}^M\Big\{
\frac{1}{2}\sum_{\lambda=\pm 1} \prod_{j=1}^k
\delta_{\lambda,\lambda_i(\ell_j,A,Z)}
\Big\}\Big\ket\!\Big\ket_{\{A,Z\}} \nonumber
\\
&=& \lim_{L\to\infty}\frac{p^{k-1}}{L^k}\sum_{\ell_1
\ldots\ell_k=1}^L \Big\bra \!\Big\bra \prod_{i=1}^M \Big\{
\prod_{j=1}^k \delta_{1,\lambda_i(\ell_j,A,Z)}
 +
\prod_{j=1}^k \delta_{-1,\lambda_i(\ell_j,A,Z)} \Big\}
 \Big\ket\!\Big\ket_{\{A,Z\}}
 \label{eq:mu_k_workedout}
\end{eqnarray}
The average $\bra\bra\ldots\ket\ket_{\{A,Z\}}$ in the last line of
(\ref{eq:mu_k_workedout}) equals the following joint probability:
\begin{eqnarray}
\hspace*{-20mm}
 \bra\bra\ldots\ket\ket_{\{A,Z\}}&=& {\rm
Prob}\left[\room ~~~~~ \Big\{
\lambda_1(\ell_1,A,Z)=\lambda_1(\ell_2,A,Z)= ~\ldots~=
        \lambda_1(\ell_k,A,Z)\Big\}
        \right.
        \nonumber\\
        \hspace*{-20mm}
&& \left.\hspace*{9mm}{\rm and}~\Big\{
\lambda_2(\ell_1,A,Z)=\lambda_2(\ell_2,A,Z)= ~\ldots~=
        \lambda_2(\ell_k,A,Z)
\Big\}
        \right.\nonumber
\\[1mm]
\hspace*{-20mm} &&\left. \hspace*{45mm}~~\vdots \right.\nonumber
\\[1mm]
\hspace*{-20mm} && \left.\hspace*{8mm}{\rm and}~ \Big\{
\lambda_{\!M}(\ell_1,A,Z)=\lambda_{\!M}(\ell_2,A,Z)= ~\ldots~=
        \lambda_{\!M}(\ell_k,A,Z)
\Big\}
        \room\right]
        \label{eq:mem_average1}
\end{eqnarray}
Let us define the short-hand
\be
{\sl
Same}(i)=\Big\{\lambda_i(\ell_1,A,Z)=\lambda_i(\ell_2,A,Z)=\ldots=\lambda_i(\ell_k,A,Z)\Big\}
\ee which states that the $i$-th component of the history string
takes the same value at the $k$ specified times
$\{\ell_1,\ldots,\ell_k\}$. Given that our bid process obeys
causality\footnote{We here use the fact that a component
$\lambda_i(\ell,A,Z)$ of the history string observed by the agents
at time $\ell$  is by construction (see definition
(\ref{eq:mem_lambda}))   referring to the overall bid at time
$\ell-i$. It follows that the probability of finding a given value
for $\lambda_i(\ell,A,Z)$ depends via causality only on the bids
at the earlier times $\{\ell-i-1,\ell-i-2,\ldots\}$, hence on
$\{\lambda_{i+1}(\ell,A,Z),\lambda_{i+2}(\ell,A,Z),\ldots\}$.},
statement (\ref{eq:mem_average1}) can be written as
\begin{eqnarray}
\bra\bra\ldots\ket\ket_{\{A,Z\}}&=& {\rm Prob}\Big[ ~{\sl
Same}(1)~\wedge~{\sl Same}(2)~\wedge~\ldots~\wedge~{\sl
Same}(M)~\Big]
  \nonumber
 \\[2mm]
 &=&~~~
{\rm Prob}\big[{\sl Same}(1)~\big|~{\sl Same}(2)\wedge \ldots
\wedge
 {\sl Same}(M)\big]\nonumber
 \\
 &&\times~
  {\rm Prob}\big[{\sl Same}(2)~\big|~{\sl Same}(3)\wedge \ldots
  \wedge
 {\sl Same}(M)\big]\nonumber
 \\[1mm]
 &&\hspace*{25mm} \vdots
 \nonumber
 \\[1mm]
  &&\times~
  {\rm Prob}\big[ {\sl Same}(M\!-\!1)~\big|~{\sl Same}(M)\big]\nonumber
 \\
  &&\times~
  {\rm Prob}\big[{\sl Same}(M)\big]
 \label{eq:mem_average2}
\end{eqnarray}
 Since we need not
consider values of $k$ which scale with $N$ or $L$, the
contributions to (\ref{eq:mu_k_workedout}) from those times
$\{\ell_1,\ldots,\ell_k\}$ for which there are correlations
between objects at a time $\ell_r$ and those at another time
$\ell_{r^\prime}$ will vanish in the limit $L\to\infty$. Since we
also know that we are in a TTI state, it follows that the
conditional probabilities in  (\ref{eq:mem_average2}) will not
depend on the actual values $\{\ell_1,\ldots,\ell_k\}$. In the
limit $L\to\infty$ we may replace \bd {\rm Prob}[{\sl
Same}(r)~\big|~{\sl Same}(r+1)\wedge \ldots
  \wedge
 {\sl Same}(M)]~\to~ {\mathcal P}_{[k|M-r]}
\ed
  where ${\mathcal P}_{[k|m]}$
 denotes the probability to find for randomly drawn and infinitely
 separated times $\{\ell_1,\ldots,\ell_k\}$ that
 $\lambda_i(\ell_1,A,Z)=\ldots=\lambda_i(\ell_k,A,Z)$,
for an index $i$, given that the identity holds for the indices
$\{i+1,\ldots,i+m\}$ (with ${\mathcal P}_{[k|0]}$ giving this
probability in the absence of conditions). This allows us to write
 (\ref{eq:mu_k_workedout}) as
\begin{eqnarray}
\mu_k &=&  p^{k-1}~ {\mathcal P}_{[k|M-1]}.{\mathcal
P}_{[k|M-2]}~\ldots~{\mathcal P}_{[k|1]}.{\mathcal P}_{[k|0]}
 \label{eq:mu_k_in_P}
\end{eqnarray}
 As a simple test one may verify (\ref{eq:mu_k_in_P})
for the trivial case $\zeta=1$ (fake history only). Here
conditioning on the past is irrelevant, so ${\mathcal
P}_{[k|m]}={\mathcal P}_{[k|0]}=2^{1-k}$ for all $m$, which indeed
gives us $\mu_k = p^{k-1}2^{(1-k)M}=1$ (as it should). In the
continuous time limit $N\to\infty$ (equivalently: for
$M\to\infty$, since $2^M=\alpha N$) we thus find the as yet exact
formula
\begin{eqnarray}
\lim_{M\to\infty}\log( \mu_k) &=&
\lim_{M\to\infty}\sum_{r=0}^{M-1} \log\Big[2^{k-1}{\mathcal
P}_{[k|r]}\Big]
 \label{eq:mu_k_in_Pagain}
\end{eqnarray}

\subsection{Reduction to history coincidence
statistics}

Next we have to find an expression for the probabilities
${\mathcal P}_{[k|r]}$. We know from
(\ref{eq:effective_bid_process},\ref{eq:finall_phistats}) that the
value of the overall bid at any time $\ell$ is only correlated
with the bid value at time $\ell^\prime$ if the two times
$(\ell,\ell^\prime)$ have {\em identical} history strings, i.e. if
$\blambda(\ell,A,Z)=\blambda(\ell^\prime,A,Z)$. We know that
individual histories show up during the process with probabilities
of order $N^{-1}$. Since the likelihood of finding {\em recurring}
histories during any number $r=\order(M)$ of consecutive
iterations of our process is thus vanishingly small (of order
$\order(M/N)$) such direct correlations are of no consequence in
our calculation. The only relevant effect of conditioning in the
sense of the ${\mathcal P}_{[k|r]}$ is via its biasing of
histories in subsequent iterations. Although the probability of
history recurrence during a time window of size $\order(M)$ is
vanishingly small, if two (short)  instances of global bid
trajectories are found to have identical realizations of some of
the bits of their history strings, they will nevertheless be more
likely than average to have an {\em identical} history realization
in the next time step. This is the subtle statistical effect
which, together with the resulting biases in the bids which are
subsequently found at times with specific histories, gives rise to
the relative history frequency distributions $\varrho(f)$ as
observed in e.g. Fig. \ref{fig:hist_frequencies}.

The statement that the conditioning in ${\mathcal P}_{[k|r]}$ acts
only via the joint likelihood of finding specific  histories
$\{\blambda_1,\ldots,\blambda_{k}\}$ at the $k$ specified (and
widely separated) times $\{\ell_1,\ldots,\ell_k\}$, translates
into
\begin{eqnarray}
{\mathcal P}_{[k|r]}&=& \sum_{\blambda_1,\ldots\blambda_k}
{\mathcal P}[k|\blambda_1,\ldots,\blambda_k] ~{\mathcal
P}[\blambda_1,\ldots,\blambda_k|r] \label{eq:historysum}
\end{eqnarray}
Here ${\mathcal P}[k|\blambda_1,\ldots,\blambda_k]$ denotes the
likelihood to find
$\blambda(\ell_1,A,Z)=\ldots=\blambda(\ell_k,A,Z)$, if the history
strings at those $k$ times equal
$\{\blambda_1,\ldots,\blambda_k\}$, and  ${\mathcal
P}[\blambda_1,\ldots,\blambda_k|r]$ denotes the likelihood of
finding those $k$ specific histories given that the bits of the
$k$ history strings have been identical over the $r$ most recent
iterations\footnote{Here one will find that consistent and
inconsistent realizations of the history noise variables
$Z(\ell,i)$ are to be treated differently: in the case of
consistent noise, one will always have
$\lambda_i(\ell,A,Z)=\lambda_{i+1}(\ell+1,A,Z)$. This is not true
for inconsistent history noise.}. The probability of finding
specific bid values $A(\ell)$ will in TTI states only depend on
the history string $\blambda$ associated with time $\ell$. Given
this history string, $A(\ell)$ is a Gaussian variable (this
follows from the effective bid process
(\ref{eq:effective_bid_process})), with some average
$\overline{A}_\blambda$ and a variance $\sigma^2_\blambda$ (which
will in due course have to be calculated). Using also the fact
that the $Z(\ell,i)$ were defined as zero average Gaussian
variables, with variance $\kappa^2$, we obtain:
\begin{eqnarray}
{\mathcal P}[k|\blambda_1,\ldots,\blambda_k]&=& \prod_{j=1}^k
\Big[\int\!DZ\int\!dA~P_{\blambda_j}(A)\theta[(1-\zeta)A+\zeta
Z]\Big] \nonumber \\ && \hspace*{5mm}+ \prod_{j=1}^k
\Big[\int\!DZ\int\!dA~P_{\blambda_j}(A)\theta[-(1-\zeta)A-\zeta
Z]\Big] \nonumber
\\
&=& \prod_{j=1}^k \left[\frac{1}{2}+\frac{1}{2}{\rm
Erf}\Big[\frac{(1-\zeta)\overline{A}_{\blambda_j}}{\sqrt{2}
\sqrt{\zeta^2\kappa^2+(1-\zeta)^2\sigma^2_{\blambda_j}}}\Big]\right]
\nonumber \\ &&\hspace*{5mm} + \prod_{j=1}^k
\left[\frac{1}{2}-\frac{1}{2}{\rm
Erf}\Big[\frac{(1-\zeta)\overline{A}_{\blambda_j}}{\sqrt{2}
\sqrt{\zeta^2\kappa^2+(1-\zeta)^2\sigma^2_{\blambda_j}}}\Big]\right]
\label{eq:Pklambdas}
\end{eqnarray}
We now write the sum over all combinations of histories in
(\ref{eq:historysum}) in terms of a partitioning in groups, where
two $M$-bit strings $\{\blambda_i,\blambda_j\}$ are in the same
 group if and only if they are identical.
We write $(g_1,g_2,\ldots)$ for the subset of all combinations
$\{\blambda_1,\ldots,\blambda_k\}$ with one group of size $g_1$, a
second group of size $g_2$, and so on\footnote{For example:
 $(k)$ denotes the subset of all
combinations $\{\blambda_1,\ldots,\blambda_k\}$ where
$\blambda_1=\ldots=\blambda_k$, $(2,1,1,\ldots)$ is the subset of
all $\{\blambda_1,\ldots,\blambda_k\}$ where precisely two history
strings are identical, and all others are distinct.}. Clearly
$g_1+g_2+\ldots=k$, for all possible subsets of our partitioning.
This allows us to write
\begin{eqnarray}
{\mathcal P}_{[k|r]}&=&
\sum_{(g_1,g_2,\ldots)}\delta_{k,g_1+g_2+\ldots} {\mathcal
P}[k|g_1,g_2,\ldots]~ {\mathcal P}[g_1,g_2,\ldots|r]
\label{eq:Pkr_again}
\end{eqnarray}
According to (\ref{eq:Pklambdas}), the distribution ${\mathcal
P}[k|g_1,g_2,\ldots]$ is of the relatively simple form
 ${\mathcal
P}[k|g_1,g_2,\ldots]= 2^{1-k}\Phi(g_1,g_2,\ldots)$, with
\begin{eqnarray}
\Phi(g_1,g_2,\ldots)&=&
 \frac{1}{2}\prod_{j\geq 1}\left\{
\sum_{\blambda}\pi_{\blambda} \left[1+{\rm
Erf}\Big[\frac{(1-\zeta)\overline{A}_{\blambda}}{\sqrt{2}
\sqrt{\zeta^2\kappa^2+(1-\zeta)^2\sigma^2_{\blambda}}}\Big]\right]^{g_j}
\nonumber\right\} \\ &&\hspace*{-5mm} + \frac{1}{2}\prod_{j\geq
1}\left\{\sum_{\blambda}\pi_{\blambda} \left[1-{\rm
Erf}\Big[\frac{(1-\zeta)\overline{A}_{\blambda}}{\sqrt{2}
\sqrt{\zeta^2\kappa^2+(1-\zeta)^2\sigma^2_{\blambda}}}\Big]\right]^{g_j}
\right\}~~  \label{eq:Pkg}
\end{eqnarray}
 Insertion of the
representation (\ref{eq:Pkr_again}) for ${\mathcal P}_{[k|r]}$
into (\ref{eq:mu_k_in_Pagain}) allows us to write the moments of
the relative history frequencies in the following form:
\begin{eqnarray}
\hspace*{-15mm} \lim_{M\to\infty}\log( \mu_k) &=&
\lim_{M\to\infty}\sum_{r=0}^{M-1} \log\left[
\sum_{(g_1,g_2,\ldots)}\delta_{k,g_1+g_2+\ldots}
\Phi(g_1,g_2,\ldots) {\mathcal P}[g_1,g_2,\ldots|r]\right]
 \label{eq:mu_k_in_Pfurther}
\end{eqnarray}
It will be helpful to assess which values of $r$ in
(\ref{eq:mu_k_in_Pfurther}) can survive the limit $M\to\infty$.
Whenever we have a value $r$ such that $M\! - \! r \!\to\! \infty$
as $M\!\to\!\infty$, the condition that the $k$ history bits were
identical over the most recent $r$ steps still leaves a large
$\order(2^{M-r})$ number of compatible history strings to be found
at the probing times $\{\ell_1,\ldots,\ell_k\}$, so  the
likelihood of finding histories coinciding in multiples
$(g_1,g_2,\ldots)$ scales as \bd {\mathcal
P}[g_1,g_2,\ldots|r]=\prod_{j|g_j>1}
\order(2^{(g_j-1)(r-M)}),~~~~~~ {\mathcal
P}[1,1,\ldots|r]=1+\order(2^{r-M})  \ed Since $k$ is finite and
$\Phi(1,1,1,\ldots)=1$, the total contribution to $\log(\mu_k)$
from those terms where $M-r\to\infty$ as $M\to\infty$ is
negligible, since for $1\ll R\ll M$ we may write
\begin{eqnarray}
&&\hspace*{-20mm} \sum_{r=0}^{R-1} \log\Big[
\sum_{(g_1,g_2,\ldots)}\delta_{k,g_1+g_2+\ldots}
\Phi(g_1,g_2,\ldots) {\mathcal P}[g_1,g_2,\ldots|r]\Big] \nonumber
\\
&=& \sum_{r=0}^{R-1} \log\Big[ 1+\order(2^{r-M})\Big]
=\order(2^{R-M})
\end{eqnarray}
Hence in (\ref{eq:mu_k_in_Pfurther}) we need only those terms
where $M-r$ is finite. These terms represent contributions where
{\em virtually all} past components of the history strings at the
times $\{\ell_1,\ldots,\ell_k\}$ were identical, which should
indeed constrain the possible histories at the times
$\{\ell_1,\ldots,\ell_k\}$ most, and indeed gives the largest
history coincidence rates. We consequently switch our conditioning
label from the number $r$ of previously identical components to
the number $M-r$ of unconstrained components, and write \bd
{\mathcal P}[g_1,g_2,\ldots|r]={\mathcal Q}[g_1,g_2,\ldots|M-r]
\ed and find (\ref{eq:mu_k_in_Pfurther}) converting into the
simpler form
\begin{eqnarray}\hspace*{-10mm}
\lim_{M\to\infty}\log( \mu_k) &=& \sum_{r\geq 1 }\log\Big[
\sum_{(g_1,g_2,\ldots)}\delta_{k,g_1+g_2+\ldots}
\Phi(g_1,g_2,\ldots) {\mathcal Q}[g_1,g_2,\ldots|r]\Big]
 \label{eq:mu_k_nearly_final}
\end{eqnarray}
We are left with the task to calculate the likelihood ${\mathcal
Q}[g_1,g_2,\ldots|r]$ of finding at the $k$ distinct times
$\{\ell_1,\ldots,\ell_k\}$ of our process the histories
$\{\blambda_1,\ldots,\blambda_k\}$ to be identical in prescribed
multiples of $(g_1,g_2,\ldots)$, given that the bits of the $k$
history vectors were identical during all but $r$ of the most
recent iterations.

At this stage  we benefit from having to consider only values of
$r$ in (\ref{eq:mu_k_nearly_final}) which are finite (compared to
$M$, which is sent to infinity). For each value $r$ of the number
of `free' components, there will be only $2^r$ possible history
strings $\blambda$ available to be allocated to the $k$ times
$\{\ell_1,\ldots,\ell_k\}$. In principle one would have to worry
about the probabilities to be assigned to each of the $2^r$
options. However, we know for the full $M$-component history
strings that their probabilities scale as $\pi_\blambda=f_\blambda
p^{-1}$ with $f_\blambda=\order(1)$, so the effective
probabilities of individual components of $\blambda\in\{-1,1\}$
must scale as \bd
\pi_{\lambda_i}=\order(\pi_{\blambda}^{1/M})=\order(\frac{1}{2}f_\blambda^{1/M})=
\frac{1}{2}[1+\order(M^{-1})] \ed From this we deduce that for
finite $r$ we may take all $2^r$ allowed history strings to have
equal probabilities. This turns the evaluation of ${\mathcal
Q}[g_1,g_2,\ldots|r]$ into a solvable combinatorial problem. Each
of $k$ elements is given randomly one of $2^r$ colours (where each
colour
 has probability $2^{-r}$), and ${\mathcal
Q}[g_1,g_2,\ldots|r]$ represents the likelihood of finding
identical colour sets of sizes $(g_1,g_2,\ldots)$. Let us
abbreviate $R=2^r$, and write the $r$-th term in
(\ref{eq:mu_k_nearly_final}) as $\log(H_r)$.  Now, using
$2^{-r(g_1+\ldots  g_R)}=2^{-rk}=R^{-k}$ we may simply
write\footnote{One easily confirms that our expression for $H_r$
is properly normalized. Upon choosing $\Phi(g_1,g_2,\ldots)=1$ one
can perform the summations iteratively, starting from $g_R$ and
descending down to $g_1$, which leads exactly to the factor $R^k$
to combine with the $R^{-k}$ present. }
\be
\lim_{M\to\infty}\log( \mu_k) = \sum_{r\geq 1 }\log H_r
 \label{eq:mu_k_verynearly_final}
\ee
\begin{eqnarray}
 \hspace*{-15mm}
H_r&=&  \sum_{(g_1,g_2,\ldots)}\Phi(g_1,g_2,\ldots) {\mathcal
Q}[g_1,g_2,\ldots|r]\nonumber
\\
\hspace*{-15mm} &=&
\sum_{g_1=0}^{k}\sum_{g_2=0}^{k-g_1}\sum_{g_3=0}^{k-g_1-g_2}
\ldots \sum_{g_{R}=0}^{k-g_1-\ldots-g_{R-1}} \Phi(g_1,g_2,\ldots)
~\delta_{k,\sum_i g_i} \nonumber
\\
\hspace*{-15mm} && \times R^{-k}
 \left(\!\begin{array}{c}k\\ g_1\end{array}\!\right)
  \!\left(\!\begin{array}{c}k - g_1\\ g_2\end{array}\!\right)
 \!\left(\!\begin{array}{c}k - g_1 - g_2\\ g_3\end{array}\!\right)
\ldots
 \left(\!\begin{array}{c}k - g_1-\ldots- g_{R-1}\\ g_R\end{array}\!\right)
\label{eq:H_r}
\end{eqnarray}

\subsection{Expansion of sign-coincidence probabilities}

Having simplified
 the
conditional distribution ${\mathcal Q}[g_1,g_2,\ldots|r]$ of
history coincidences, we turn to $\Phi(g_1,g_2,\ldots)$
 as given by  (\ref{eq:Pkg}).
 If we restrict ourselves to an expansion of  (\ref{eq:Pkg})
 in powers of the (random) bid biases
 $\overline{A}_{\blambda}$ in which we retain only the leading
 terms, our problem simplifies further to the point where we can
 obtain a fully explicit expression for the moments $\mu_k$.
 In \ref{app:expandingphi} we derive the following
compact relations:
 \begin{eqnarray}
 \Phi(1,1,1,\ldots)&=& 1\label{eq:trivial_phi}
 \\
 \Phi(g_1,g_2,\ldots)&=& e^{\frac{1}{2}\Omega \sum_{j\geq
 1}g_j(g_j-1)-\frac{1}{4}\Omega^2 \sum_j g_j
 (g_j-1)(2g_j-3)+\order(\Omega^3)}
 \label{eq:expansion_of_phi}
 \\
 \Omega
 &=& \sum_{\blambda}\pi_{\blambda}~
{\rm Erf}^2\Big[\frac{(1-\zeta)\overline{A}_{\blambda}}{\sqrt{2}
\sqrt{\zeta^2\kappa^2+(1-\zeta)^2\sigma^2_{\blambda}}}\Big]
\label{eq:define_Omega_mem}
\end{eqnarray}
The results (\ref{eq:trivial_phi},\ref{eq:expansion_of_phi}) imply
that, rather than knowing the full probability distribution
${\mathcal P}[g_1,g_2,\ldots|r]$ in (\ref{eq:mu_k_in_Pfurther}),
we only need the (conditional) statistics of a modest number of
relatively simple monomials. Expanding the exponential in
(\ref{eq:expansion_of_phi}) up to the relevant orders, and using
$\sum_j g_j=k$ (which is always true inside (\ref{eq:H_r}))
produces
\begin{eqnarray}
\Phi(g_1,g_2,\ldots)&=& 1+\frac{1}{2}\Omega \Big[\sum_{j\geq
 1}g^2_j-k\Big]\label{eq:phi_inter1}
 \\
 &&\hspace*{-15mm}+\frac{1}{4}\Omega^2\Big[
 \frac{1}{2}\sum_{ij\geq 1}g_i^2g_j^2
- 2\sum_{j\geq 1} g_j^3 -(k-5)\sum_{j\geq 1} g_j^2
+\frac{1}{2}k^2\!-3k
 \Big]
 +\order(\Omega^3)
 \nonumber
\end{eqnarray}
Since the combinatorial averaging process of (\ref{eq:H_r}) in
this particular representation involves a measure which is
invariant under permutations of the numbers $\{g_1,g_2,\ldots\}$,
 the average of (\ref{eq:phi_inter1})  is identical to that of
the following simpler function (with $R=2^r$):
\begin{eqnarray}
\Phi_{\rm eff}(g_1,g_2,\ldots)&=& 1+\frac{1}{2}\Omega (R
g_1^2-k)\label{eq:phi_inter2}
 \\
 &&\hspace*{-29mm}+\frac{1}{8}\Omega^2\Big[
 R g_1^4
+ R(R-1)g_1^2g_2^2 - 4R g_1^3 -2(k-5)R g_1^2 +k^2\!-6k
 \Big]
 +\order(\Omega^3)
 \nonumber
\end{eqnarray}
Instead of having to use full combinatorial measure
(\ref{eq:H_r}),  we can therefore extract all the relevant
information from the (joint) marginal distribution for the pair
$(g_1,g_2)$ only. Inserting (\ref{eq:phi_inter2}) into
(\ref{eq:H_r}) gives us
\begin{eqnarray}
H_r&=& 1+\frac{1}{2}\Omega \Big[ R G_{2,0}^{k,R} -k\Big]
+\frac{1}{8}\Omega^2\Big[
 R G_{4,0}^{k,R}
+ R(R-1) G_{2,2}^{k,R} \nonumber
 \\
 &&\hspace*{20mm}
 - 4R G_{3,0}^{k,R} -2(k-5)R G_{2,0}^{k,R}
+k^2\!-6k
 \Big]
 +\order(\Omega^3)~~~~~
\label{eq:H_r_infactors}
\end{eqnarray}
with
\begin{eqnarray}
G_{a,b}^{k,R}&=&
\sum_{g_1=0}^{k}\sum_{g_2=0}^{k-g_1}\sum_{g_3=0}^{k-g_1-g_2}
\ldots \sum_{g_{R}=0}^{k-g_1-\ldots-g_{R-1}} g_1^a g_2^b
~\delta_{k,\sum_i g_i} \nonumber
\\
&& \hspace*{-3mm}
\times R^{-k}
 \left(\!\begin{array}{c}k\\ g_1\end{array}\!\right)
  \!\left(\!\begin{array}{c}k - g_1\\ g_2\end{array}\!\right)
 \!\left(\!\begin{array}{c}k - g_1 - g_2\\ g_3\end{array}\!\right)
\ldots
 \left(\!\begin{array}{c}k - g_1-\ldots- g_{R-1}\\ g_R\end{array}\!\right)
 \nonumber
\\
&=& R^{-k}\sum_{g_1=0}^{k}\sum_{g_2=0}^{k-g_1}
\left(\!\begin{array}{c}k\\ g_1\end{array}\!\right)
  \!\left(\!\begin{array}{c}k - g_1\\ g_2\end{array}\!\right)
  (R-2)^{k-g_1-g_2}g_1^a g_2^b
  \label{eq:Gfactors}
\end{eqnarray}
Those combinatorial factors $G_{a,b}^{k,R}$ which we need in order
to evaluate (\ref{eq:H_r_infactors}) are calculated in
\ref{app:memorycombinatorics}. They are found to be
\begin{eqnarray*}
G_{2,0}^{k,R}&=& \frac{k}{R} + \frac{k(k-1)}{R^{2}}
\\
G_{3,0}^{k,R} &=&  \frac{k}{R}   + \frac{3k(k-1)}{R^2}
  + \frac{k(k-1)(k-2)}{R^{3}}
\\
G_{4,0}^{k,R}&=&
 \frac{k}{R}  +  \frac{7k(k-1)}{R^{2}}
 + \frac{6k(k-1)(k-2)}{R^{3}}
 + \frac{k(k-1)(k-2)(k-3)}{R^{4}}
 \\
G_{2,2}^{k,R}
  &=& \frac{k(k-1)}{R^{2}}   +   \frac{2k(k-1)(k-2)}{R^{3}}
  + \frac{k(k-1)(k-2)(k-3)}{R^{4}}
\end{eqnarray*}
Insertion of these factors into (\ref{eq:H_r_infactors}), followed
by restoration of the short-hand $R=2^r$,  gives us the fully
explicit expression
\begin{eqnarray}
H_r&=& 1+\frac{1}{2}\Omega  k(k-1)2^{-r} +\frac{1}{8}\Omega^2
k(k-1)(k-2)(k-3) 4^{-r}
 +\order(\Omega^3)~~~~~~
\label{eq:H_r_explicit}
\end{eqnarray}
We can now write explicit formulae for the moments of the relative
history frequencies, and hence also for the distribution
$\varrho(f)$ itself, in the form  an expansion in a parameter
$\Omega$ which controls the width of this distribution.

\subsection{Resulting prediction for $\varrho(f)$}

\begin{figure}[t]
\vspace*{1mm} \hspace*{20mm} \setlength{\unitlength}{0.91mm}
\begin{picture}(90,65)
\put(2,5){\epsfysize=60\unitlength\epsfbox{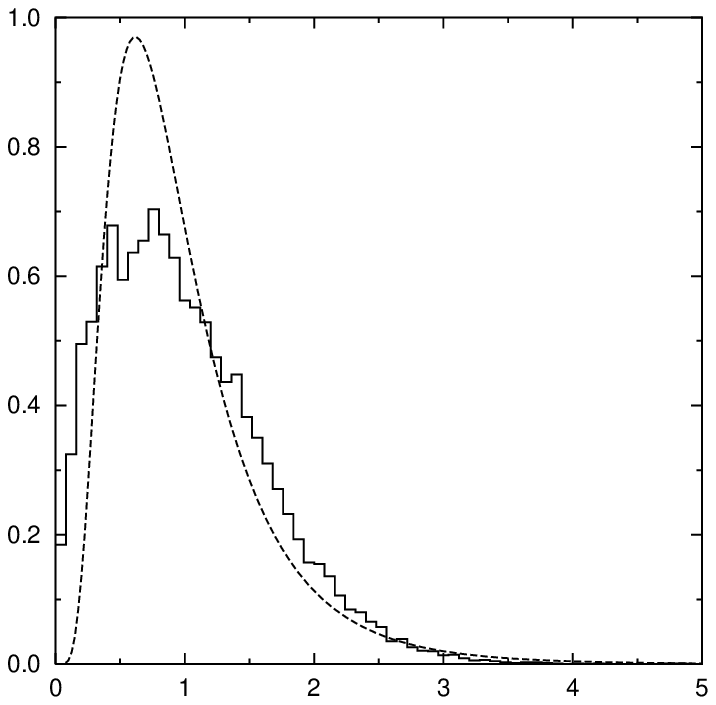}}
\put(37,0){\here{$\freq$}}  \put(-3,35){\here{$\varrho(\freq)$}}

 \put(82,5){\epsfysize=60\unitlength\epsfbox{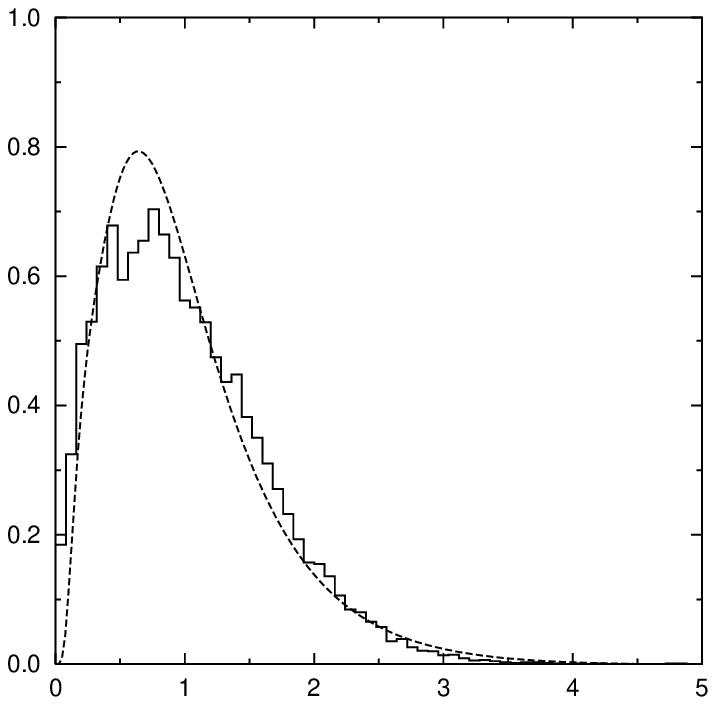}}
\put(117,0){\here{$\freq$}}  \put(77,35){\here{$\varrho(\freq)$}}

\end{picture}
\vspace*{3mm} \caption{Test of the predicted history frequency
distributions (\ref{eq:rhof_order1}) (left picture, based on
expansion of the moments $\mu_k$ up to first order in the width,
$\mu_k=e^{\frac{1}{2}\Omega  k(k-1)}$) and (\ref{eq:rhof_order2})
(right picture, based on expansion  up to second order,
$\mu_k=e^{\frac{1}{2}\Omega  k(k-1)
 -\frac{1}{12}\Omega^2 k(k-1)(2k-3)}$),
together with the data of Fig. \ref{fig:hist_frequencies} as
measured in simulations for $\alpha=2.0$ and $N=8193$. In both
cases the second moment which parametrizes  (\ref{eq:rhof_order1})
and (\ref{eq:rhof_order2}) was taken from the data: $\mu_2\approx
1.380$.  } \label{fig:hist_frequency_theory}
\end{figure}

The result (\ref{eq:H_r_explicit}), together with the earlier
relation (\ref{eq:mu_k_verynearly_final}) and the geometric series
leads us finally to the desired expression for the
moments\index{moments of history frequency distribution} $\mu_k$:
\begin{eqnarray}
\lim_{M\to\infty}\log( \mu_k) &=&
 \frac{1}{2}\Omega  k(k-1)
 -\frac{1}{12}\Omega^2 k(k-1)(2k-3)
 +\order(\Omega^3)
 \label{eq:final_muk}
\end{eqnarray}
We see that this general formula obeys $\mu_0=\mu_1=1$, as it
should, and that
\be
\lim_{M\to\infty}\mu_2 = e^{\Omega
 -\frac{1}{6}\Omega^2
 +\order(\Omega^3)}
 \label{eq:final_mu2}
\ee
 Insertion into our earlier expression
(\ref{eq:rho_f}) for $\varrho(f)$ leads in the limit $M\to\infty$
to a formula in which, at least up the relevant orders in
$\Omega$, the insertion of a Gaussian integral allows us to carry
out the summation over moments explicitly:
\begin{eqnarray}
\hspace*{-5mm} \varrho(f)&=&  \int\!\frac{d\omega}{2\pi}e^{i\omega
f} \sum_{k\geq 0}\frac{ (-i\omega)^k}{k!} e^{ \frac{1}{2}\Omega
k(k-1)
 -\frac{1}{12}\Omega^2 k(k-1)(2k-3)
 +\order(\Omega^3)}
 \nonumber
 \\
 \hspace*{-5mm}
 &=&\int\!Dz  \int\!\frac{d\omega}{2\pi}e^{i\omega f} \sum_{k\geq
0}\frac{ (-i\omega)^k}{k!} \Big[ 1
 -\frac{1}{6}\sqrt{\Omega}\frac{d^3}{dz^3}
 +\ldots\Big]e^{ zk\sqrt{\Omega
+\frac{5}{6}\Omega^2} -\frac{1}{2}k(\Omega +\frac{1}{2}\Omega^2) }
 \nonumber
 \\
 \hspace*{-5mm}
 &=&\int\!Dz \Big[ 1
 +\frac{1}{6}\sqrt{\Omega}(3z-z^3)
 +\ldots\Big]~\delta\Big[ f- e^{ z\sqrt{\Omega
+\frac{5}{6}\Omega^2} -\frac{1}{2}(\Omega +\frac{1}{2}\Omega^2) }
\Big] \label{eq:rhof_in_gaussian}
\end{eqnarray}
We may use (\ref{eq:final_mu2}) to express $\Omega$ in terms of
$\mu_2$, turning our expansion of the moments $\mu_k$ into an
expansion in powers of $\log(\mu_2)$. Depending on whether we wish
to take our expansion only to order $\order(\log(\mu_2))$, or also
to $\order(\log^2(\mu_2))$, we obtain
\begin{eqnarray}
{\rm to}~\order(\log(\mu_2)): &~~~& \varrho(f) =
\frac{e^{-\frac{1}{2}z^2(f)}}
 {f\sqrt{2\pi\log(\mu_2)}}
\label{eq:rhof_order1}
 \\
 &&
 z(f)=\frac{\log(f)+\frac{1}{2}\log(\mu_2)}{\sqrt{\log(\mu_2)}}
\\
{\rm to}~\order(\log^2(\mu_2)): &~~~& \varrho(f) =
\frac{e^{-\frac{1}{2}z^2(f)}\Big[ 1
 +\frac{1}{6}\sqrt{\log(\mu_2)}\big(3z(f)-z^3(f)\big)\Big]}
 {f\sqrt{2\pi [\log(\mu_2)+\log^2(\mu_2)]}}~~~~~~
 \label{eq:rhof_order2}\\
 &&
 z(f)=\frac{\log(f)+\frac{1}{2}\big[\log(\mu_2)+\frac{2}{3}\log^2(\mu_2)\big]}{\sqrt{\log(\mu_2)+\log^2(\mu_2)}}
\end{eqnarray}
The two statements (\ref{eq:rhof_order1}) and
(\ref{eq:rhof_order2}) are indeed found to constitute increasingly
accurate predictions for the actual distribution of the relative
history frequencies, see e.g. Fig.
\ref{fig:hist_frequency_theory}.   We have thus been able to
explain the origin and the characteristics of the observed history
frequency statistics. However, both formulae are expansions for
small $\Omega$. Should (\ref{eq:rhof_order2}) be applied to values
of $\Omega$ which are not small, one has to be careful in dealing
with large values of $f$, where $\varrho(f)$ could become negative
(this would have been prevented by the higher orders in $\Omega$).
The implication is that in the Gaussian integral
(\ref{eq:rhof_in_gaussian}) one must in practice either introduce
a cut-off $z_c=\order(\Omega^{-1/6})$, or exponentiate the factor
$[ 1
 +\frac{1}{6}\sqrt{\log(\mu_2)}\big(3z(f)-z^3(f)\big)]$.

\subsection{The width of $\varrho(f)$}

What remains in order to round off our analysis of the
distribution of relative history frequencies is to calculate the
width parameter $\Omega$ in (\ref{eq:final_muk}) self-consistently
from our equations. According to our theory, $\Omega$ is given by
(\ref{eq:define_Omega_mem}), i.e. by
\begin{eqnarray}
\Omega
 &=& \sum_{\blambda}\pi_{\blambda}~
{\rm Erf}^2\Big[\frac{(1-\zeta)\overline{A}_{\blambda}}{\sqrt{2}
\sqrt{\zeta^2\kappa^2+(1-\zeta)^2\sigma^2_{\blambda}}}\Big]
\label{eq:define_Omega_mem_again}
\end{eqnarray}
The quantities $\overline{A}_\blambda$ and
$\sigma^2_\blambda=\overline{A^2}_{\blambda}-\overline{A}_{\blambda}$
describe the statistics of those bids which correspond to times
with a prescribed history string $\blambda$. We know from
(\ref{eq:Aexplicit}) that these are Gaussian variables, which
implies that $\overline{A}_\blambda$ and $\sigma^2_{\blambda}$ are
all we need to know. Since we restrict ourselves to non-anomalous
TTI states, we can write both as long-time averages:
\begin{eqnarray}
\overline{A}_{\blambda}&=& \pi_{\blambda}^{-1}\lim_{L\to\infty}
L^{-1}\sum_{\ell=1}^L \delta_{\blambda,\blambda(\ell,A,Z)} A(\ell)
\\
\overline{A^2}_{\blambda}&=&\pi_{\blambda}^{-1}
\lim_{L\to\infty}L^{-1}\sum_{\ell=1}^L
\delta_{\blambda,\blambda(\ell,A,Z)} A^2(\ell)
\end{eqnarray}
We can work out the average $\overline{A}_{\blambda}$, using
(\ref{eq:Aexplicit}) and time-translation invariance, and
subsequently  define the new time variables
$s_i=\ell_i-\ell_{i+1}$ (for $i<r$) and $s_r=\ell_r$ (so that
$\ell_j=s_j+s_{j+1}+\ldots+s_r$). This results in
\begin{eqnarray}
\overline{A}_{\blambda}&=& \lim_{L\to\infty}
\frac{1}{Lp\pi_{\blambda}}\sum_{\ell_0=1}^L \sum_{r\geq
0}(-\del)^r\sum_{\ell_1\ldots \ell_r} G(\ell_0\!-\ell_1)\ldots
G(\ell_{r-1}\!-\ell_r) \nonumber \\ &&\hspace*{20mm}\times~
\Big[\prod_{i=0}^r p\delta_{\blambda,\blambda(\ell_i,A,Z)} \Big]
\phi_{\ell_r}\nonumber \\
 &=&
\frac{1}{p\pi_{\blambda}}\sum_{r\geq 0}(-\del)^r \sum_{s_0 \ldots
s_{r-1}}G(s_0)\ldots G(s_{r-1}) \nonumber
\\
&&\hspace*{10mm} \lim_{L\to\infty}
\frac{1}{L}\sum_{s_r=0}^{L-\sum_{i<r}s_i} \Big[\prod_{i=0}^r
p\delta_{\blambda,\blambda(s_i+\ldots+s_r,A,Z)} \Big] \phi_{s_r}
\end{eqnarray}
Given our ansatz of short history correlation times, in the sense
of  (\ref{eq:history_corr_time}), and given
$\chi=\sum_{\ell>0}G(\ell)<\infty$ (so $G(\ell)$ must decay
sufficiently fast), we find this expression simplifying to
\begin{eqnarray}
\overline{A}_{\blambda}&=& \sum_{r\geq 0}(-\chi p\pi_{\blambda})^r
\lim_{L\to\infty} \frac{1}{\pi_{\blambda}L}\sum_{s=0}^{L}
\delta_{\blambda,\blambda(s,A,Z)}\phi_{s}
=\frac{\overline{\phi}_{\blambda}}{1+\chi p\pi_{\blambda}}
\label{eq:Aconditional}
\end{eqnarray}
In a similar manner we find
\begin{eqnarray}
\overline{A^2}_{\blambda}&=&
\lim_{L\to\infty}\frac{1}{\pi_{\blambda}L}\sum_{\ell_0\ell_0^\prime=0}^{L}\delta_{\blambda,\blambda(\ell_0,A,Z)}
 A(\ell_0)A(\ell^\prime_0) \delta_{\ell_0\ell^\prime_0}\nonumber
\\
&=&
\lim_{L\to\infty}\frac{1}{\pi_{\blambda}L}\sum_{\ell_0\ell_0^\prime=0}^{L}
\sum_{r,r^\prime\geq 0}(-\del)^{r+r^\prime} \sum_{\ell_1\ldots
\ell_r} G(\ell_0\!-\ell_1)\ldots G(\ell_{r-1}\!-\ell_r) \nonumber
\\
&&\times \sum_{\ell^\prime_1\ldots \ell^\prime_r}
G(\ell^\prime_0\!-\ell^\prime_1)\ldots
G(\ell^\prime_{r^\prime-1}\!-\ell^\prime_{r^\prime})
\delta_{\blambda,\blambda(\ell_0,A,Z)}  \nonumber
\\
&&\times~ \Big[\prod_{i=1}^{r}
p\delta_{\blambda,\blambda(\ell_i,A,Z)} \Big]
\Big[\prod_{i=1}^{r^\prime}
p\delta_{\blambda,\blambda(\ell^\prime_i,A,Z)} \Big]
\delta_{\ell_0\ell^\prime_0} \phi_{\ell_r}
\phi_{\ell^\prime_{r^\prime}}
 \nonumber
 \\
&=& \frac{1}{p\pi_{\blambda}} \sum_{r,r^\prime\geq
0}(-\del)^{r+r^\prime} \!\sum_{s_0 \ldots s_{r-1}}\!G(s_0)\ldots
G(s_{r-1}) \!\sum_{s^\prime_0 \ldots
s^\prime_{r^\prime-1}}\!G(s^\prime_0)\ldots
G(s^\prime_{r^\prime-1}) \nonumber
\\
&&\times~
\lim_{L\to\infty}\frac{1}{L}\sum_{s_r=0}^{L-\sum_{i<r}s_i}
\sum_{s^\prime_{r^\prime}=0}^{L-\sum_{i<r^\prime}s^\prime_i}
p\delta_{\blambda,\blambda(s_0+\ldots+s_r,A,Z)}
  \nonumber
\\
&&\times~\Big[\prod_{i=1}^r
p\delta_{\blambda,\blambda(s_i+\ldots+s_r,A,Z)}
\Big]\Big[\prod_{i=1}^{r^\prime}
p\delta_{\blambda,\blambda(s^\prime_i+\ldots+s^\prime_{r^\prime},A,Z)}
\Big] \delta_{\sum_i s_i,\sum_i s_i^\prime} \phi_{s_r}
\phi_{s^\prime_{r^\prime}}
 \nonumber\\
 &&\label{eq:nasty_variance}
\end{eqnarray}
 Again we use $\sum_{\ell}G(\ell)<\infty$
to justify that in the summations over $s_r$ and
$s^\prime_{r^\prime}$ the upper limit can safely be replaced by
$L$. Thus:
\begin{eqnarray}
\overline{A^2}_{\blambda}&=&
 \frac{1}{p\pi_{\blambda}} \sum_{r,r^\prime\geq
0}(-\del)^{r+r^\prime} \!\sum_{s_0 \ldots s_{r-1}}\!G(s_0)\ldots
G(s_{r-1}) \!\sum_{s^\prime_0 \ldots
s^\prime_{r^\prime-1}}\!G(s^\prime_0)\ldots
G(s^\prime_{r^\prime-1}) \nonumber
\\
&&\times~ \lim_{L\to\infty}\frac{1}{L}\sum_{s_r=0}^{L}
p\delta_{\blambda,\blambda(\sum_j s_j,A,Z)} \Big[\prod_{i=1}^r
p\delta_{\blambda,\blambda(\sum_{j\geq i} s_j,A,Z)} \Big]
  \nonumber
\\
&&\times~\Big[\prod_{i=1}^{r^\prime}
p\delta_{\blambda,\blambda(\sum_j s_j-\sum_{j<i}s_j^\prime,A,Z)}
\Big] \phi_{s_r} \phi_{\sum_{j}s_j-\sum_{j<r^\prime}s_j^\prime}
\end{eqnarray}
The present calculation is similar to that of the volatility
matrix in the fake history online MG \cite{CoolHeim01} (the
quantity
$\sigma^2_{\blambda}=\overline{A^2}_{\blambda}-\overline{A}^2_{\blambda}$
can be regarded as a conditional volatility, where the condition
is that in collecting our statistics we are to restrict ourselves
to those times where  the observed history strings take the value
$\blambda$), so also here we have to worry about pairwise time
coincidences. Each such coincidence effectively removes one
constraint of the type $\delta_{\blambda,\blambda(\ldots,A,Z)}$,
since the latter will be met automatically. The remaining terms
will occur in extensive summations, so that we may replace each
`unpaired' occurrence of a factor
$\delta_{\blambda,\blambda(\ldots,A,Z)}$, except for those with
the same argument as one of the Gaussian variables $\phi$, by its
time average $\pi_{\blambda}$. In practice this implies the
replacement
\begin{eqnarray}
&&\Big[\prod_{i=1}^{r-1} p\delta_{\blambda,\blambda(\sum_{j\geq i}
s_j,A,Z)} \Big]\Big[\prod_{i=1}^{r^\prime-1}
p\delta_{\blambda,\blambda(\sum_j s_j-\sum_{j<i}s_j^\prime,A,Z)}
\Big]~\rightarrow~\hspace*{20mm}\nonumber \\ && \hspace*{20mm} ~~~
(p\pi_{\blambda})^{r+r^\prime-2}
\pi_{\blambda}^{-\sum_{i=1}^{r-1}\sum_{j=1}^{r^\prime-1}\delta_{\sum_{\ell\geq
i} s_\ell,\sum_\ell s_\ell-\sum_{\ell<j}s_\ell^\prime}} \nonumber
\\
&&\hspace*{20mm} =
(p\pi_{\blambda})^{r+r^\prime-2}\prod_{i=1}^{r-1}\prod_{j=1}^{r^\prime-1}\Big[1+
\frac{1-\pi_{\blambda}}{p\pi_{\blambda}}\frac{\rate}{2\del}\delta_{\sum_{\ell<i}
s_\ell,\sum_{\ell<j}s_\ell^\prime}\Big]
\end{eqnarray}
and therefore
\begin{eqnarray}
\hspace*{-20mm} \overline{A^2}_{\blambda}&=&
  \sum_{r,r^\prime\geq
0}(-\del)^{r+r^\prime} \!\sum_{s_0 \ldots s_{r-1}}\!G(s_0)\ldots
G(s_{r-1}) \!\sum_{s^\prime_0 \ldots
s^\prime_{r^\prime-1}}\!G(s^\prime_0)\ldots
G(s^\prime_{r^\prime-1}) \nonumber
\\
\hspace*{-20mm}
&&\times(p\pi_{\blambda})^{r+r^\prime}\prod_{i=1}^{r-1}\prod_{j=1}^{r^\prime-1}\Big[1+
\frac{1-\pi_{\blambda}}{p\pi_{\blambda}}\frac{\rate}{2\del}\delta_{\sum_{\ell<i}
s_\ell,\sum_{\ell<j}s_\ell^\prime}\Big]
  \nonumber
\\
\hspace*{-20mm}
&&\times\sum_{k}\delta_{k,\sum_{j<r}s_j-\sum_{j<r^\prime}s_j^\prime}
\lim_{L\to\infty}\frac{1}{\pi^2_{\blambda}L}\sum_{s=0}^{L}
 \delta_{\blambda,\blambda(s,A,Z)}
\delta_{\blambda,\blambda(s+k,A,Z)}  \phi_{s} \phi_{s+k}
\end{eqnarray}
As in the calculation of the volatility in \cite{CoolHeim01},
lacking as yet a method to deal with all the complicated terms
generated by the factor proportional to the learning rate $\rate$,
we have to restrict ourselves in practice to approximations. As in
\cite{CoolHeim01}  we first remove the most tricky terms by
putting $\rate\to 0$. This gives
\begin{eqnarray}
\hspace*{-20mm} \overline{A^2}_{\blambda}&=&
  \sum_{r,r^\prime\geq
0}(-\del p\pi_{\blambda})^{r+r^\prime} \!\sum_{s_0 \ldots
s_{r-1}}\!G(s_0)\ldots G(s_{r-1}) \!\sum_{s^\prime_0 \ldots
s^\prime_{r^\prime-1}}\!G(s^\prime_0)\ldots
G(s^\prime_{r^\prime-1}) \nonumber
\\
\hspace*{-20mm}
&&\times~\sum_{k}\delta_{k,\sum_{j<r}s_j-\sum_{j<r^\prime}s_j^\prime}
\lim_{L\to\infty}\frac{1}{\pi^2_{\blambda}L}\sum_{s=0}^{L}
 \delta_{\blambda,\blambda(s,A,Z)}
\delta_{\blambda,\blambda(s+k,A,Z)}  \phi_{s} \phi_{s+k}
\end{eqnarray}
We then assume that the limit $L\to\infty$ in the last line
converts the associated sample average into a full average over
the statistics of the Gaussian fields $\phi_\ell$ given by
(\ref{eq:finall_phistats}), i.e.
\begin{eqnarray*}
&& \hspace*{-5mm}\lim_{L\to\infty}
\frac{1}{\pi^2_{\blambda}L}\sum_{s=0}^{L}
 \delta_{\blambda,\blambda(s,A,Z)}
\delta_{\blambda,\blambda(s+k,A,Z)}  \phi_{s} \phi_{s+k} ~\to~
\nonumber
\\
&& \hspace*{5mm}
\lim_{L\to\infty}\frac{1}{\pi^2_{\blambda}L}\sum_{s=0}^{L}
 \delta_{\blambda,\blambda(s,A,Z)}
\delta_{\blambda,\blambda(s+k,A,Z)}\bra \phi_{s}
\phi_{s+k}\ket_{\phi|A,Z} = \frac{1}{2}[1+C(k)]
\end{eqnarray*}
Separating the correlation function into a persistent and a
non-persistent term, $C(k)=c+\tilde{C}(k)$, and returning to the
earlier notation with time differences inside the kernels $G$,
results in the history-conditioned equivalent of the volatility
approximation in \cite{CoolHeim01}:
\begin{eqnarray}
\hspace*{-10mm} \overline{A^2}_{\blambda}&=&
  \sum_{r,r^\prime\geq
0}(-\del p\pi_{\blambda})^{r+r^\prime} \!\sum_{s_0 \ldots
s_{r-1}}\!G(s_0)\ldots G(s_{r-1}) \!\sum_{s^\prime_0 \ldots
s^\prime_{r^\prime-1}}\!G(s^\prime_0)\ldots
G(s^\prime_{r^\prime-1}) \nonumber
\\
\hspace*{-10mm} && \times~
\frac{1}{2}\Big[1+c+\tilde{C}\big(\sum_{j<r}s_j-\sum_{j<r^\prime}s_j^\prime\big)\Big]
 \nonumber\\
 \hspace*{-10mm}
 &=&
 \frac{1+c}{2(1+\chi p\pi_{\blambda})^2}
 +\frac{1}{2}\int\!ds
 ds^\prime ~(\one+p\pi_{\blambda}G)^{-1}\tilde{C}(s-s^\prime)(\one+p\pi_{\blambda}G)^{-1}
\end{eqnarray}
where $\one(x,y)=\delta(x-y)$. In order to get to the present
stage we have averaged the $\phi$-dependent terms inside
$\overline{A^2}_{\blambda}$ over the Gaussian measure $\bra
\ldots\ket_{\phi|A,Z}$. Consistency demands that in working out
$\sigma^2_{\blambda}=\overline{A^2}_{\blambda}-\overline{A}^2_{\blambda}$
we do the same with the term $\overline{A}^2_{\blambda}$, where
$\overline{A}_{\blambda}$ is given by (\ref{eq:Aconditional}), so
 our approximation for the history-conditioned volatility
becomes
\begin{eqnarray}
\bsigma^2_{\blambda}&=& \overline{A^2}_{\blambda}-
\frac{\bra\overline{\phi}^2_{\blambda}\ket}{(1+\chi
p\pi_{\blambda})^2} \nonumber
\\
&=& \overline{A^2}_{\blambda}- \lim_{L\to\infty}\frac{1}{(L
\pi_{\blambda})^{2}}\sum_{\ell\ell^\prime=1}^L\delta_{\blambda,\blambda(\ell,A,Z)}
\delta_{\blambda,\blambda(\ell^\prime,A,Z)}
~\frac{1+C(\ell-\ell^\prime)}{2(1+\chi p\pi_{\blambda})^2}
\nonumber
\\
&=& \frac{1}{2}\int\!ds
 ds^\prime
 ~(\one+p\pi_{\blambda}G)^{-1}(s)\tilde{C}(s-s^\prime)(\one+p\pi_{\blambda}G)^{-1}(s^\prime)
 \label{eq:sigmalambda_nearlythere}
\end{eqnarray}
Our final step again follows \cite{CoolHeim01}. We assume that the
  non-persistent correlations $\tilde{C}(t)$ decay vary fast, away from the value $\tilde{C}(0)=1-c$,
  so that in the expansion of
 (\ref{eq:sigmalambda_nearlythere}) in powers of $G$ we retain
 only the zero-th term:
 \begin{eqnarray}
\bsigma^2_{\blambda} &=& \frac{1}{2}(1-c)
 \label{eq:sigmalambda_final}
\end{eqnarray}
We may now return to expression (\ref{eq:define_Omega_mem_again})
 and insert our approximations (\ref{eq:Aconditional}) and
(\ref{eq:sigmalambda_final}):
\begin{eqnarray}
\Omega
 &=&\lim_{p\to\infty} \sum_{\blambda}\pi_{\blambda}~
{\rm
Erf}^2\left[\frac{(1-\zeta)\overline{\phi}_{\blambda}}{\sqrt{2}(1+\chi
p\pi_{\blambda})
\sqrt{\zeta^2\kappa^2+\frac{1}{2}(1-\zeta)^2(1-c)}}\right]
\nonumber
\\
&=&  \int\!df d\phi~\varrho(f,\phi) f~ {\rm
Erf}^2\left[\frac{(1-\zeta)\phi}{\sqrt{2}(1+\chi f)
\sqrt{\zeta^2\kappa^2+\frac{1}{2}(1-\zeta)^2(1-c)}}\right]~~~
\label{eq:nearly_found_Omega}
\end{eqnarray}
with
\be
\varrho(f,\phi)=\lim_{p\to\infty}\frac{1}{p}\sum_{\blambda}\delta[f-p\pi_{\blambda}]\delta[\phi-\overline{\phi}_{\blambda}]
\ee
 We know the $\overline{\phi}_{\blambda}$ to be Gaussian
variables, with $\bra \overline{\phi}_{\blambda}\ket=0$ and $\bra
\overline{\phi}^2_{\blambda}\ket=\frac{1}{2}(1+c)$ (see the above
derivation of $\sigma^2_{\blambda}$ where this was shown and
used). Hence, upon making our final simplifying assumption that in
the relevant orders of our calculation the correlations between
the history frequencies $\pi_{\blambda}$ and the Gaussian fields
$\overline{\phi}_{\blambda}$ are irrelevant, we obtain
\be
\varrho(f,\phi)=\varrho(f)~
\frac{e^{-\phi^2/(1+c)}}{\sqrt{\pi(1+c)}} \ee and hence
(\ref{eq:nearly_found_Omega}) simplifies to
\begin{eqnarray}
\Omega &=&  \int_0^\infty\!df~\varrho(f) f \int\!Dx~{\rm
Erf}^2\left[\frac{x(1-\zeta)\sqrt{1+c}}{2(1+\chi f)
\sqrt{\zeta^2\kappa^2+\frac{1}{2}(1-\zeta)^2(1-c)}}\right]~~~
\label{eq:found_Omega_nearly}
\end{eqnarray}
Using the integral $\int\!Dx~{\rm
Erf}^2(Ax)=\frac{4}{\pi}\arctan[\sqrt{1+4A^2}]-1$, in combination
with the identity $\int\!df~\varrho(f)f=1$, our approximate
expression for the parameter  $\Omega$ thus becomes
\begin{eqnarray}
\hspace*{-15mm} \Omega &=&
\frac{4}{\pi}\int_0^\infty\!df~\varrho(f) f~ \arctan\left[1+
\frac{(1-\zeta)^2(1+c)}{(1+\chi f)^2
[\zeta^2\kappa^2+\frac{1}{2}(1-\zeta)^2(1-c)]}\right]^{\frac{1}{2}}
-1 \label{eq:found_Omega}
\end{eqnarray}
In the limit of strictly fake history we recover from
(\ref{eq:found_Omega}) the value $\lim_{\zeta\to
1}\Omega=(4/\pi)\arctan[1]-1=0$, as it should. For MGs with
strictly true market history, on the other hand, expression
(\ref{eq:found_Omega}) simplifies to
\begin{eqnarray}
\lim_{\zeta\to 0}\Omega &=&
\frac{4}{\pi}\int_0^\infty\!df~\varrho(f) f~ \arctan\left[1+
\frac{2(1+c)}{(1+\chi f)^2 (1-c)}\right]^{\frac{1}{2}} -1
\label{eq:found_Omega_memonly}
\end{eqnarray}
In accordance with earlier observations in simulations
\cite{ChalletMarsili00} we also see that, as the system approaches
the phase transition when $\alpha$ is lowered from within the
ergodic regime, the increase  of the susceptibility $\chi$
automatically reduces the width parameter $\Omega$, until it
vanishes completely at the critical point.

\section{Closed theory for persistent observables in the ergodic regime}

We have now obtained a closed theory for the time-translation
invariant states of our MG, albeit in approximation. It consists
of the equations
(\ref{eq:mem_closed1},\ref{eq:mem_closed2},\ref{eq:mem_closed3},\ref{eq:mem_closed4})
for the persistent order parameters, combined with expressions
(\ref{eq:rhof_order1},\ref{eq:rhof_order2}) for the shape and
(\ref{eq:final_mu2},\ref{eq:found_Omega}) for the width of the
relative history frequency distribution $\varrho(f)$. This theory
predicts correctly (i) that the phase transition point
$\alpha_c(T)$ of the MG with history is identical to that of the
model with fake memory, (ii) that at the transition point the
relative history frequency distribution reduces to
$\varrho(\freq)=\delta[\freq-1]$ (with at that point also the
order parameters all becoming independent of whether we have true
or fake history), and (iii) the shape of the relative history
frequency distribution. In the limit $\alpha\to \infty$ the theory
also reproduces the correct order parameter values
$\chi=\phi=c=0$, for any value of $\zeta$. For $\zeta=0$ (strictly
true memory) it predicts
$\lim_{\alpha\to\infty}\Omega=\frac{1}{3}$ and hence
$\lim_{\alpha\to\infty} \mu_2\approx 1.37$. \vsp

\begin{figure}[t]
\vspace*{1mm} \hspace*{20mm} \setlength{\unitlength}{0.95mm}
\begin{picture}(160,65)
\put(0,5){\epsfysize=60\unitlength\epsfbox{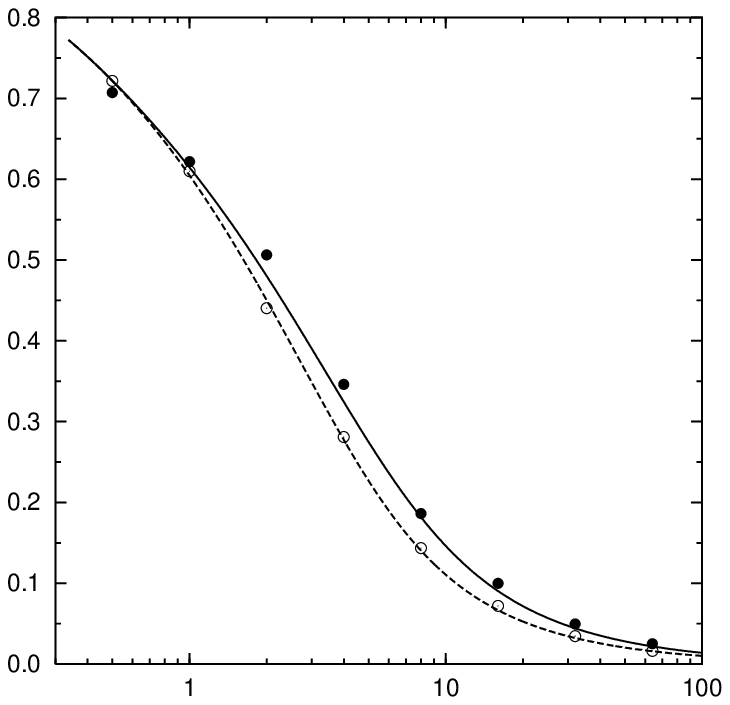}}
\put(35,0){\here{$\alpha$}}  \put(-1,35){\here{$c$}}

\put(78,5){\epsfysize=60\unitlength\epsfbox{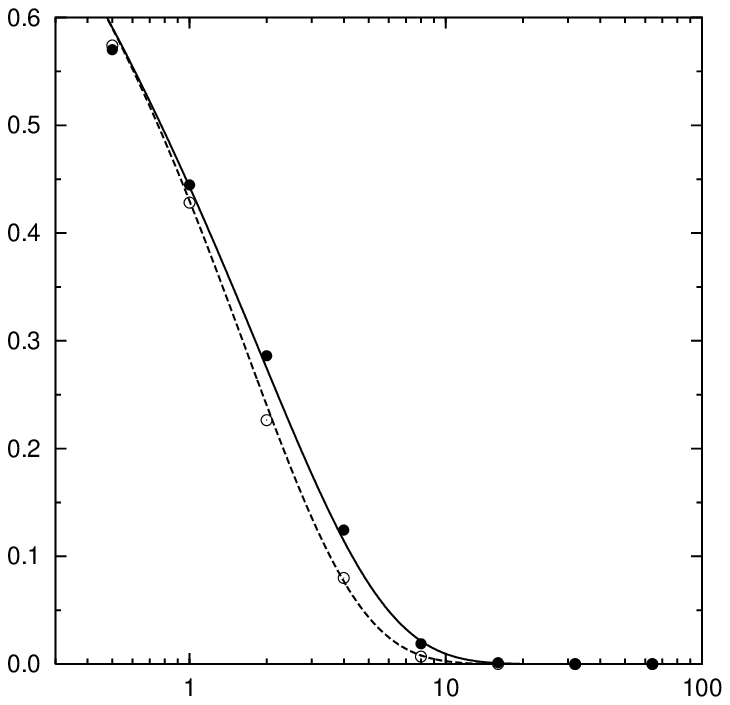}}
\put(113,0){\here{$\alpha$}}  \put(77,35){\here{$\phi$}}

\end{picture}
\vspace*{-3mm} \caption{Left: the predicted persistent
correlations $c$  together with simulation data in the non-ergodic
regime, for the on-line MG with strictly true history (i.e.
$\zeta=0$; the solid line gives the theoretical prediction, full
circles the experimental data) and for the on-line MG with
strictly fake memory (i.e. $\zeta=1$; the dashed line gives the
theoretical prediction, open circles the experimental data). In
both cases decision noise was absent. Right: the corresponding
predicted fraction $\phi$ of frozen agents, under the same
experimental conditions and with the same meaning of lines and
markers.} \label{fig:memory_cphi}
\end{figure}

\begin{figure}[t]
\vspace*{4mm} \hspace*{50mm} \setlength{\unitlength}{0.95mm}
\begin{picture}(90,65)
\put(0,5){\epsfysize=60\unitlength\epsfbox{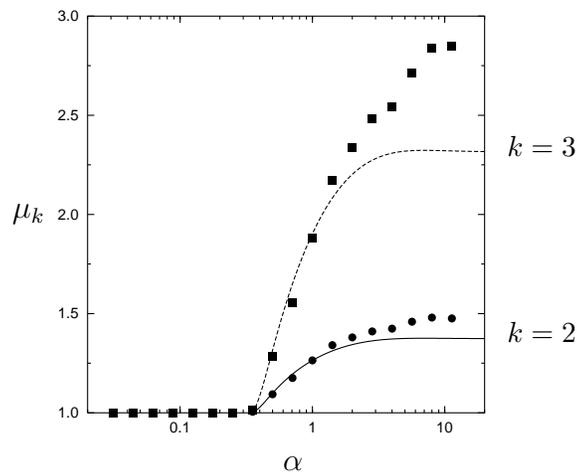}}
\put(35,0){\here{$\alpha$}}  \put(-2,35){\here{$\mu_k$}}
\put(65,43){\small $k=3$} \put(65,17){\small $k=2$}

\end{picture}
\vspace*{2mm} \caption{The moments $\mu_2=\int\!df~\varrho(f)f^2$
and  $\mu_3=\int\!df~\varrho(f)f^3$ of the distribution of
relative history frequencies for the MG with strictly true history
and absent decision noise (i.e. $\zeta=T=0$), as predicted by the
theory (solid and dashed lines), compared to the moments as
measured in numerical simulations (markers, with circles
indicating  $\mu_2$ and squares indicating $\mu_3$). Note that
$\mu_0=\mu_1=1$ (by definition). } \label{fig:memory_moments}
\end{figure}

Let us finally reduce our closed equations to a more compact form,
for the simplest nontrivial case of the MG with strictly true
market history (i.e. $\zeta=0$) and without decision noise (i.e.
$\sigma[\infty]=1$). Here we have
\begin{eqnarray}
u&=&\frac{\sqrt{\alpha}\chiR}{S_0\sqrt{2}}~~~~~~~~\chi=
\frac{1-\phi}{\alpha \chiR}~~~~~~~~\phi =1- {\rm Erf}[u]
\\
c &=& 1-{\rm Erf}[u]+\frac{1}{2u^2}{\rm
Erf}[u]-\frac{1}{u\sqrt{\pi}}e^{-u^2} \label{eq:mem_c}
\\
\chiR&=&\int\!Dz \Big[ 1
 +\frac{1}{6}\sqrt{\Omega}(3z-z^3)
 \Big] \Big[e^{ -z\sqrt{\Omega
+\frac{5}{6}\Omega^2} +\frac{1}{2}(\Omega +\frac{1}{2}\Omega^2) }+
\chi\Big]^{-1} \label{eq:final_overview_chiR}
\\
S_0^2&=& (1+c)\int\!Dz \Big[ 1
 +\frac{1}{6}\sqrt{\Omega}(3z-z^3)
 \Big]\Big[
 e^{- z\sqrt{\Omega
+\frac{5}{6}\Omega^2} +\frac{1}{2}(\Omega +\frac{1}{2}\Omega^2)
}+\chi \Big]^{-2} \label{eq:final_overview_S0}
\\
\Omega &=& \int\!Dz \Big[ 1
 +\frac{1}{6}\sqrt{\Omega}(3z-z^3)
 \Big] e^{ z\sqrt{\Omega
+\frac{5}{6}\Omega^2} -\frac{1}{2}(\Omega +\frac{1}{2}\Omega^2)
}\nonumber
\\
&& \times~\left\{ \frac{4}{\pi}\arctan\Big[1+
\frac{2(1+c)}{(1-c)\Big[1+\chi e^{ z\sqrt{\Omega
+\frac{5}{6}\Omega^2} -\frac{1}{2}(\Omega +\frac{1}{2}\Omega^2)
}\Big]^2 }\Big]^{\frac{1}{2}}\!\! -1\right\}~~~~~~
\label{eq:final_overview_Omega}
\end{eqnarray}
Upon using (\ref{eq:mem_c}) to write $c$ as a function of $u$,
i.e. $c=c(u)$ with $c(u)$ denoting the right-hand side of
(\ref{eq:mem_c}), and upon eliminating the quantities $\phi$ and
$S_0$, we find ourselves with a closed set of equations for the
trio  $\{u,\chi,\Omega\}$:
\begin{eqnarray}
u&=&\frac{{\rm Erf}[u]}{\chi\sqrt{2\alpha(1+c)}} \left\{\int\!Dz
\frac{1
 +\frac{1}{6}\sqrt{\Omega}(3z-z^3)}{\Big[
 e^{- z\sqrt{\Omega
+\frac{5}{6}\Omega^2} +\frac{1}{2}(\Omega +\frac{1}{2}\Omega^2)
}+\chi \Big]^{2}} \right\}^{-\frac{1}{2}} \label{eq:mem_closed_u}
\\[1mm]
\chi&=& \frac{{\rm Erf}[u]}{\alpha} \left\{\int\!Dz \frac{1
 +\frac{1}{6}\sqrt{\Omega}(3z-z^3)}{e^{ -z\sqrt{\Omega
+\frac{5}{6}\Omega^2} +\frac{1}{2}(\Omega +\frac{1}{2}\Omega^2) }+
\chi}\right\}^{-1} \label{eq:mem_closed_chi}
\\[1mm]
\Omega &=& \int\!Dz \Big[ 1
 +\frac{1}{6}\sqrt{\Omega}(3z-z^3)
 \Big] e^{ z\sqrt{\Omega
+\frac{5}{6}\Omega^2} -\frac{1}{2}(\Omega +\frac{1}{2}\Omega^2)
}\label{eq:mem_closed_Omega}
\\
&& \times\left\{ \frac{4}{\pi}\arctan\Big[1+
\frac{2[1+c(u)]}{[1-c(u)]\Big[1+\chi e^{ z\sqrt{\Omega
+\frac{5}{6}\Omega^2} -\frac{1}{2}(\Omega +\frac{1}{2}\Omega^2)
}\Big]^2 }\Big]^{\frac{1}{2}}\!\! -1\right\}\nonumber
\end{eqnarray}

Solving these three coupled equations numerically, followed by
comparison with simulation data, shows a surprising level of
agreement, in spite of the expansions and assumptions which have
been used to derive
 (\ref{eq:mem_closed_u},\ref{eq:mem_closed_chi},\ref{eq:mem_closed_Omega}).
Figure \ref{fig:memory_cphi} shows the performance of the theory
in describing the on-line MG with strictly true market history
(i.e. $\zeta=0$), together with similar data for the on-line fake
history MG (i.e. $\zeta=1$), for comparison\footnote{Below the
critical point, where $\chi=\infty$ throughout, equation
(\ref{eq:found_Omega}) predicts that $\Omega=0$. This implies that
$\varrho(f)=\delta[f-1]$ for $\alpha<\alpha_c(T)$, and that below
the critical point the differences between true and fake history
(if any) are confined to dynamical phenomena or to states without
time-translation invariance. This confirms earlier observations in
numerical simulations \cite{ChalletMarsili00}, where it was found
that the persistent order parameters in MGs with and without
history were identical in the low $\alpha$ regime.}. In all these
simulations $N=8193$. Calculation of the first two non-trivial
moments $\mu_k$ of the distribution of relative history
frequencies, see e.g. figure \ref{fig:memory_moments} (where in
the simulations $\alpha N^2=2^{28}$), shows that for small values
of the width of $\varrho(f)$ (i.e. $\mu_2$ close to one, which is
true close to and below the critical point) the predictions of the
theory are excellent, but that the performance of equations
(\ref{eq:mem_closed_u},\ref{eq:mem_closed_chi},\ref{eq:mem_closed_Omega})
deteriorates for larger values of $\mu_2$. This is obvious, since
these equations result effectively from an expansion for small
values of $\mu_2-1$. Taking this expansion to higher orders
 should lead to systematic improvement, but will
be non-trivial.

\section{Discussion}

We have developed a
 mathematical procedure for the derivation of
exact dynamical solutions for Minority Games with {\em real}
market histories, using the generating functional analysis
techniques of \cite{DeDominicis}. So far these techniques had only
been developed for (and applied successfully to) the less
realistic but mathematically simpler MG versions with fake market
histories, restricting theoretical progress to those particular
game versions only. We have shown how the technical difficulties
associated with the non-Markovian character of the microscopic
laws induced by having real histories can be dealt with, and found
(in the infinite system size limit) exact and closed macroscopic
laws from which to solve the canonical dynamic order parameters
for the standard (on-line) MG with true market history. Here these
laws turn out to be formulated in terms of {\em two} effective
equations (rather than a single equation, as for models with fake
histories): one for an effective agent, and one for an effective
overall market bid. In the second part of this paper we have
constructed solutions for these effective equations, focusing
mostly on the usual persistent observables of the MG in
time-translation invariant states (persistent correlations and the
fraction of frozen agents) and on the calculation from first
principles of the distribution of history frequencies. These
objects are calculated in the form of an expansion in powers of
the width of the history frequency distribution, of which the
first few terms are derived in explicit form. The final theory was
shown to give accurate predictions for the persistent observables
and for the shape of the history frequency distribution. It gives
precise predictions for the width in the region where this width
remains relatively small (which is inevitable in view of the
expansion used).

\appendix

\section{Recovering the fake history limit}
\label{app:fake_only}

It  helpful for our understanding of the $N\to\infty$ limit in
(\ref{eq:finallR},\ref{eq:finallSigma}) to first return to the
simplest case where we know what the outcome should be, being
$\zeta=1$, i.e. fake history strings of the inconsistent type
(\ref{eq:typeB}). This is the model which was solved in
\cite{CoolHeim01}. In doing so we {\em en passant} re-confirm the
correctness of the assumed scaling $\del=\rate/2p$.

For $\zeta=1$ we  see in (\ref{eq:mem_lambda}) and
(\ref{eq:overlineW}) that both $\blambda(\ldots)$ and
$\overline{W}[\ldots]$ lose their dependence on the path $\{A\}$,
and reduce to
\be
\blambda(\ell,Z)=\left(\begin{array}{c} \sgn[Z(\ell,1)]\\
 \vdots
\\
\sgn[ Z(\ell,M)]
\end{array}\right)~~~~~~~~
\overline{W}[\ell,\ell^\prime;Z] =
\delta_{\blambda(\ell,Z),\blambda(\ell^\prime,Z)}
\label{eq:zetazero} \ee
 The role of
the Gaussian variables $\{Z\}$ has thereby been reduced to
determining the statistics of the symmetric random matrix $\B$
with entries $\B_{\ell
\ell^\prime}=\overline{W}[\ell,\ell^\prime;Z]$:
\begin{eqnarray}
 \cP[\B]=\bigbra
\prod_{\ell,\ell^\prime}\delta\Big[
\B_{\ell\ell^\prime}-\prod_{\lambda=1}^{\log_2(p)}\theta\big[Z(\ell,\lambda)Z(\ell^\prime,\lambda)\big]\Big]\bigket_{\!\!\{Z\}}
 \label{eq:Wnomem_LU}
\end{eqnarray}
with $p=2^M=\alpha N$. The two relevant properties of these
matrices are relatively easily derived, and are found to be the
following. For any cyclic combination of $r$-th moments (with
$r>0$), where $s_1>s_2>\ldots
>s_r$ and $r>1$ (no summations) one has
\begin{eqnarray}
\bra\B_{s_1 s_2}\B_{s_2,s_3}\ldots \B_{s_r,s_1}\ket_{\B} &=&
p^{1-r}
 \label{eq:Bmoments_LUR}
\end{eqnarray}
The second type of average one needs involves two time-ordered
strings of
 matrix elements (of lengths $r$ and $r^\prime$,
respectively) connected by two further matrix elements, where
$s_0>s_1>\ldots
>s_r$ and $s^\prime_0>s^\prime_1>\ldots
>s^\prime_{r^\prime}$:
\begin{eqnarray}
\bra [ \B_{s_0 s_1}\B_{s_1 s_2} \ldots \B_{s_{r-1} s_r}]\B_{s_r
s^\prime_{r^\prime}}[ \B_{s_0^\prime s^\prime_1}\B_{s^\prime_1
s^\prime_2}\ldots \B_{s^\prime_{r^\prime-1} s^\prime_{r^\prime}}]
\B_{s_0 s^\prime_0} \ket &&\nonumber\\ && \hspace*{-50mm}=
p^{\sum_{i=0}^{r}\sum_{j=0}^{r^\prime}\delta_{s_i
s^\prime_j}-r-r^\prime-1} \label{eq:Bmoments_LUsig}
\end{eqnarray}
 The partial
decoupling of the paths $\{A\}$ and $\{Z\}$ implies that our
expressions for the kernels $R$ and $\Sigma$ simplify to
\begin{eqnarray}
\room R(t,t^\prime) &=& \lim_{\del\to 0}~ \frac{\delta}{\delta
A_e(t^\prime)} ~\left. \bigbra \B_{\ell\ell^\prime} \bigbra
A(\ell) \bigket_{\!\{A|\B\}}
 \bigket_{\B}\right|_{\ell=t/\del,\ell^\prime=t^\prime/\del}~~~~
 \label{eq:Rzetazero}
 \\
 \room
\Sigma(t,t^\prime) &=& \rate ~\lim_{\del\to 0} ~\frac{1}{\del}
\left.\bigbra \B_{\ell\ell^\prime} \bigbra
A(\ell)A(\ell^\prime)\bigket_{\!\{A|\B\}} \bigket_{\B}
\right|_{\ell=t/\del,\ell^\prime=t^\prime/\del}~~~~
 \label{eq:Sigmazero}
\end{eqnarray}
Since the bid evolution process (\ref{eq:Astats}) is now linear in
$\{A\}$, and involves only $\{A\}$-independent zero-average
Gaussian fields $\phi_\ell$, where $\bra
\phi_\ell\phi_{\ell^\prime}\ket_{\{\phi|\B\}}=\frac{1}{2}\B_{\ell\ell^\prime}[1+C(\ell,\ell^\prime)]$,
it is easily solved for any given realization of the random matrix
$\B$:
\begin{eqnarray}
A(\ell)=A_e(\ell)+\phi_\ell+\sum_{r>0}(-\frac{\rate}{2})^r
\sum_{k<\ell}[(G\!\B)^r]_{\ell k}~[ A_e(k)+\phi_k]
\label{eq:Asolution}
\end{eqnarray}
in which $G\!\B$ denotes the matrix with entries
$(G\!\B)_{\ell\ell^\prime}=G(\ell,\ell^\prime)\B_{\ell\ell^\prime}$
(i.e. involving component multiplication rather than matrix
multiplication). To make a comparison with the results of
\cite{CoolHeim01} we must remove the external bid perturbations
$A_e(\ell)$ after they have served to generate the response
function $R$.

We can evaluate (\ref{eq:Rzetazero}) using only expression
(\ref{eq:Asolution}), the causality of the response function,  and
formula (\ref{eq:Bmoments_LUR}). These give, with $\delta/\delta
A_e(\ell)=\del^{-1}\!\partial/\partial A(\ell)$:
\begin{eqnarray}
R(\ell,\ell^\prime) &=&\lim_{A_e\to 0} \lim_{\del\to
0}\frac{\partial}{\partial A_e(\ell^\prime)} \frac{1}{\del}
\int\!d\B~\cP[\B] ~\B_{\ell\ell^\prime} ~\bra
 A(\ell)
 \ket_{\{\phi|\B\}}
 \nonumber
\\
&=& \lim_{\del\to 0}  \frac{1}{\del} \sum_{r\geq
0}(-\frac{\rate}{2})^r \int\!d\B~\cP[\B] ~\B_{\ell\ell^\prime}
[(G\!\B)^r]_{\ell \ell^\prime }  \nonumber
\\
&=& \lim_{\del\to 0}  \frac{1}{\del}\left\{\room
\delta_{\ell\ell^\prime} - \del G(\ell,\ell^\prime) \nonumber
\right.
\\
&&\left.\hspace*{10mm}
 +
 \sum_{r> 1}(-\del)^r
\! \sum_{s_2>s_3>\ldots >s_{r}}\!\!
 G(\ell,s_2) G(s_2,s_3)\ldots
 G(s_{r},\ell^\prime)
 \right\}
 \label{eq:nomemR}
\end{eqnarray}
We observe in (\ref{eq:nomemR}), in view of
$\delta_{\ell\ell^\prime}\to \delta_N\delta(t-t^\prime)$ in the
limit  $N\to\infty$, that the canonical scaling of time (modulo
$\order(1)$ factors) is indeed $\del=\rate/2p$. We then find
exactly the expression in \cite{CoolHeim01}  for the on-line `fake
history' MG:
\begin{eqnarray}
R(t,t^\prime)
 &=& \delta(t-t^\prime)+
\sum_{r> 0}(-1)^r G^r(t,t^\prime) =[\one+ G]^{-1}(t,t^\prime)
 \label{eq:nomemRnew}
\end{eqnarray}
Had we chosen an alternative scaling with $N$ of $\delta_N$, we
would have found either the trivial result $R=0$, or an
ill-defined expression.

 Next we turn to expression (\ref{eq:finallSigma}) for the
 effective agent's noise covariances, with $A_e=0$.
 The equivalence of the present expression and that in \cite{CoolHeim01}
 will be more transparent upon renaming
$(\ell,\ell^\prime)\to(s_0,s_0^\prime)$ and
$D(k,k^\prime)=1+C(k,k^\prime)$:
 \begin{eqnarray}
\Sigma(s_0,s_0^\prime) &=& \lim_{\del\to 0} ~\frac{\rate}{\del}
\int\!d\B~\cP[\B] ~\B_{s_0 s_0^\prime}~\bra A(s_0)A(s_0^\prime)
\ket_{\{\phi|\B\}} \nonumber
\\
 &=&
\lim_{\del\to 0} \frac{\rate}{2\del} \sum_{r,r^\prime\geq
0}(-\frac{\rate}{2})^{r+r^\prime}\sum_{s_1\ldots
s_r}\sum_{s_1^\prime\ldots
s^\prime_{r^\prime}}D(s_r,s^\prime_{r^\prime})\nonumber
\\
&& \times~ G(s_0,s_1)\ldots
G(s_{r-1},s_r)~G(s_0^\prime,s_1^\prime)\ldots
G(s^\prime_{r^\prime-1},s^\prime_{r^\prime})\nonumber
\\[1mm]
&&\times ~\bra (\B_{s_0 s_1} \ldots \B_{s_{r-1} s_r}) \B_{s_r
s^\prime_{r^\prime}} (\B_{s^\prime_0 s^\prime_1} \ldots
\B_{s^\prime_{r-1} s^\prime_r}) \B_{s_0 s^\prime_{0}} \ket_{\B}
~~~~~~~~ \label{eq:nonmem_generalsigma}
\end{eqnarray}
with the proviso that when $r=0$ we must interpret the sums as
$\sum_{s_1\ldots s_2} \to 1$, $G(s_0,s_1)\ldots G(s_{r-1},s_r)\to
1$ and $\B_{s_0 s_1} \ldots \B_{s_{r-1} s_r}\to 1$ (and similarly
when $r^\prime=0$). Since the kernel $\Sigma(s_0,s^\prime_0)$ is
symmetric, we may without loss of generality choose
$s^\prime_0\geq s_0$. Dependent on the whether any or both of the
indices $(r,r^\prime)$ are zero, we have to evaluate the following
averages (with the short-hand $\notdelta_{ij}=1-\delta_{ij}$):
\begin{itemize}
\item $r=r^\prime=0$: here the average of the last line in (\ref{eq:nonmem_generalsigma})
reduces to
\begin{eqnarray}
\bra \ldots\ket_{\B}&=& \bra \B^2_{s_0 s_0^\prime}\ket=\bra
\B_{s_0 s_0^\prime}\ket =\delta_{s_0
s^\prime_0}+\frac{1}{p}\notdelta_{s_0 s_0^\prime}
\end{eqnarray}
\item $r^\prime=0,~r>0$: here the average in
(\ref{eq:nonmem_generalsigma}) reduces to two terms (representing
the cases $s_0=s_0^\prime$ versus $s_0<s^\prime_0$), which are
both of the form
 (\ref{eq:Bmoments_LUR}),
\begin{eqnarray}
\bra \ldots\ket_{\B}&=& \bra (\B_{s_0 s_1} \ldots \B_{s_{r-1}
s_r}) \B_{s_r s^\prime_{0}}  \B_{s_{0} s^\prime_{0}} \ket =
p^{-r}\delta_{s_{0} s^\prime_0}+ p^{-r-1}\notdelta_{s_0
s^\prime_0}~~~~~
\end{eqnarray}
where we used $\B_{kk}=1$, for any $k$. The case
$r=0$,~$r^\prime>0$ is clearly equivalent.
 \vspace*{1mm}
\item $r,r^\prime>0$: now the relevant average reduces to that of
(\ref{eq:Bmoments_LUsig}),
\begin{eqnarray}
\bra \ldots \ket_{\B}&=& \bra [ \B_{s_0 s_1}\ldots \B_{s_{r-1}
s_r}]\B_{s_r s^\prime_{r^\prime}}[ \B_{s_0^\prime
s^\prime_1}\ldots \B_{s^\prime_{r^\prime-1}
s^\prime_{r^\prime}}]\B_{s_0 s^\prime_0} \ket\nonumber \\
 &=&
p^{\sum_{i=0}^{r}\sum_{j=0}^{r^\prime}\delta_{s_i
s^\prime_j}-r-r^\prime-1} \label{eq:nomem_LUnasty}
\end{eqnarray}
\end{itemize}
Expression (\ref{eq:nomem_LUnasty}) reduces to those derived for
the cases where $r$ or $r^\prime$ is zero (or both), so it is true
for any  $(r,r^\prime)$. We may thus insert
(\ref{eq:nomem_LUnasty}) into (\ref{eq:nonmem_generalsigma}), and
obtain:
 \begin{eqnarray}
\Sigma(s_0,s_0^\prime) &=& \lim_{\del\to 0} \sum_{r,r^\prime\geq
0}(-\del)^{r+r^\prime}\sum_{s_1\ldots s_r}\sum_{s_1^\prime\ldots
s^\prime_{r^\prime}}D(s_r,s^\prime_{r^\prime})
\prod_{i=0}^{r}\prod_{j=0}^{r^\prime}\Big[1+(p-1)\delta_{s_i
s^\prime_j}\Big] \nonumber
\\
&& \hspace*{10mm}\times~ G(s_0,s_1)\ldots
G(s_{r-1},s_r)~G(s_0^\prime,s_1^\prime)\ldots
G(s^\prime_{r^\prime-1},s^\prime_{r^\prime}) \nonumber
\\
&=& \lim_{N\to\infty} \sum_{r,r^\prime\geq
0}(-\del)^{r+r^\prime}\sum_{s_1\ldots s_r}\sum_{s_1^\prime\ldots
s^\prime_{r^\prime}}D(s_r,s^\prime_{r^\prime}) \nonumber
\\
&&\hspace*{10mm}\times
\prod_{i=0}^{r}\prod_{j=0}^{r^\prime}\Big[1+\frac{\rate}{2\del}\delta_{s_i
s^\prime_j}[1-\order(\del)]\Big] \nonumber
\\
&& \hspace*{10mm}\times~ G(s_0,s_1)\ldots
G(s_{r-1},s_r)~G(s_0^\prime,s_1^\prime)\ldots
G(s^\prime_{r^\prime-1},s^\prime_{r^\prime}) \nonumber
\\
&=& \sum_{r,r^\prime\geq
0}(-1)^{r+r^\prime}\int_0^\infty\!ds_1\ldots
 ds_r ds_1^\prime\ldots
 ds^\prime_{r^\prime}
\prod_{i=0}^{r}\prod_{j=0}^{r^\prime}\Big[1+\frac{1}{2}
\rate\delta[s_i- s^\prime_j]\Big] \nonumber
\\
&& \hspace*{10mm} \times~ G(s_0,s_1)\ldots
G(s_{r-1},s_r)~G(s_0^\prime,s_1^\prime)\ldots
G(s^\prime_{r^\prime-1},s^\prime_{r^\prime}) \nonumber
\\
&& \hspace*{10mm} \times~[1+ C(s_r,s^\prime_{r^\prime})]
\label{eq:nomemSigmanew}
\end{eqnarray}
Also (\ref{eq:nomemSigmanew}) is identical to the corresponding
expression in \cite{CoolHeim01},  as it should.

\section{Expansion of bid sign recurrence probabilities}
\label{app:expandingphi}

Here we derive the expansion (\ref{eq:expansion_of_phi}) of the
function $\Phi(g_1,g_2,\ldots)$ as defined in (\ref{eq:Pkg}). We
abbreviate
\be
E_\blambda={\rm
Erf}\Big[\frac{(1-\zeta)\overline{A}_{\blambda}}{\sqrt{2}
\sqrt{\zeta^2\kappa^2+(1-\zeta)^2\sigma^2_{\blambda}}}\Big]
\label{eq:Elambda}
 \ee
with $\sum_{\blambda}\pi_\blambda E^{r}_{\blambda}=\bra
 E^r\ket$. These short-hands allow us to compactify (\ref{eq:Pkg}) to
\begin{eqnarray}
\Phi(g_1,g_2,\ldots)&=&
 \frac{1}{2}\prod_{j\geq 1}\Big[
\sum_{\blambda}\pi_{\blambda} (1+E_{\blambda})^{g_j} \Big] +
\frac{1}{2}\prod_{j\geq 1}\Big[\sum_{\blambda}\pi_{\blambda}
(1-E_{\blambda})^{g_j}\Big]
 \nonumber
\\
&=&
 \frac{1}{2}\prod_{j\geq
 1}\Big[\sum_{n=0}^{g_j}
 \left(\!\begin{array}{c} g_j\\n\end{array}\!\right)
 \bra E^n\ket \Big] + \frac{1}{2}\prod_{j\geq
1}\Big[ \sum_{n=0}^{g_j}
 \left(\!\begin{array}{c} g_j\\n\end{array}\!\right)
 (-1)^n
 \bra E^n\ket \Big]
 \nonumber
 \\
&=&
 \sum_{n_1=0}^{g_1}\sum_{n_2=0}^{g_2}\ldots
 \left(\!\begin{array}{c} g_1\\n_1\end{array}\!\right)\!
 \left(\!\begin{array}{c} g_2\\n_2\end{array}\!\right)
\ldots
 \frac{1}{2}[1+(-1)^{n_1+n_2+\ldots}]
 \bra E^{n_1}\ket  \bra E^{n_2}\ket \ldots
 \nonumber\\&&
\end{eqnarray}
Since the overall average bid in the MG is equally likely to be
positive than negative, and since (\ref{eq:Elambda}) tells us that
$\sgn[E_\blambda]=\sgn[\overline{A}_\blambda]$, the moments $\bra
E^r\ket$ for even values of $r$ will have to be zero. From this it
follows that
\begin{eqnarray}
\Phi(g_1,g_2,\ldots)&=&
 \sum_{0\leq n_1\leq \frac{1}{2}g_1}\sum_{0\leq n_2\leq \frac{1}{2}g_2}\ldots
 \left(\!\begin{array}{c} g_1\\ 2n_1\end{array}\!\right)\!
 \left(\!\begin{array}{c} g_2\\ 2n_2\end{array}\!\right)
\ldots
 \bra E^{2n_1}\ket  \bra E^{2n_2}\ket \ldots
 \nonumber\\
 &=&\prod_{j\geq 1}\left[1+
\sum_{1\leq n\leq g_j/2}
 \left(\!\begin{array}{c} g_j\\ 2n\end{array}\!\right)
 \bra E^{2n}\ket\right]
 \nonumber
 \end{eqnarray}
 so
 \begin{eqnarray}
\log \Phi(g_1,g_2,\ldots)&=& \sum_{j\geq 1}\log\Big[
 1+\sum_{1\leq n \leq g_j/2}
 \left(\!\begin{array}{c} g_j\\ 2n\end{array}\!\right)\!
 \bra E^{2n}\ket\Big]
 \label{eq:phi_ex}
 \end{eqnarray}
 Equation (\ref{eq:phi_ex}) tells us, firstly, that
 \be
\Phi(1,1,1,\ldots)=1 \label{eq:phi111}
 \ee
For arbitrary history coincidence numbers $(g_1,g_2,\ldots)$, not
necessarily all equal to one, we may expand (\ref{eq:phi_ex}) in
the moments $\bra E^r\ket$:
\begin{eqnarray}
\hspace*{-25mm} \log\Phi(g_1,g_2,\ldots)
 &=&
\sum_{j\geq 1}\log\Big[
 1+\frac{1}{2}g_j(g_j-\! 1) \bra E^{2}\ket
+\frac{1}{24}g_j(g_j-\! 1)(g_j-\! 2)(g_j-\! 3)
 \bra E^{4}\ket
 \nonumber
 \\
 \hspace*{-25mm}
 &&\hspace*{80mm}
+\order(\bra E^6\ket)
 \Big]\nonumber
 \\
 \hspace*{-25mm}
  &&
\hspace*{-25mm}=  \frac{1}{2} \sum_{j\geq 1} g_j(g_j-\! 1) \left\{
 \room\bra E^{2}\ket
 +\frac{1}{12}\Big[ (g_j-\! 2)(g_j-\! 3)
 \bra E^{4}\ket
- 3g_j(g_j-\! 1)\bra E^{2}\ket^2\Big]
 \right\}
 +\order(\bra E^6\ket)
 \nonumber
\end{eqnarray}
Finally, in leading order in $E$ we may regard the variables
$E_\blambda$ as proportional to $\overline{A}_\blambda$, and
therefore as distributed in a Gaussian manner. This implies (since
$\bra E\ket=0$) that in leading order we have $\bra E^4\ket=3\bra
E^2\ket$. Hence
\begin{eqnarray}
\log\Phi(g_1,g_2,\ldots)&=&
  \frac{1}{2}\bra E^{2}\ket
\sum_{j\geq 1} g_j(g_j-1)
 -\frac{1}{4}\bra E^{2}\ket^2
\sum_{j\geq 1} g_j(g_j-1)(2g_j-3) \nonumber
\\
&&\hspace*{50mm}
 +\order(\bra E^6\ket)
\end{eqnarray}

\section{Combinatorics in history frequency moments}
\label{app:memorycombinatorics}

In this appendix we calculate the combinatorial factors
$G_{a,b}^{k,R}$ as defined in (\ref{eq:Gfactors}). They can be
obtained by differentiation of a simple generating function:
\begin{eqnarray}
G_{a,b}^{k,R} &=& R^{-k}\sum_{g_1=0}^{k}\sum_{g_2=0}^{k-g_1}
\left(\!\begin{array}{c}k\\ g_1\end{array}\!\right)
  \!\left(\!\begin{array}{c}k - g_1\\ g_2\end{array}\!\right)
  (R-2)^{k-g_1-g_2}g_1^a g_2^b
  \nonumber
  \\
  &=&
  R^{-k}\lim_{x,y\to 1}(x\frac{d}{dx})^a
  (y\frac{d}{dy})^b(R-2+x+y)^{k}
\end{eqnarray}
In particular:
\begin{eqnarray}
\hspace*{-25mm}
 G_{2,0}^{k,R}&=& kR^{-1} + k(k-1)R^{-2}
\\[1mm]
\hspace*{-25mm} G_{3,0}^{k,R}&=&  kR^{-1}   + 3k(k-1)R^{-2}
  + k(k-1)(k-2)R^{-3}
\\[1mm]
\hspace*{-25mm} G_{4,0}^{k,R}&=&
 k R^{-1}  +  7k(k-1)R^{-2}
 + 6k(k-1)(k-2)R^{-3}
 + k(k-1)(k-2)(k-3)R^{-4}
 \\[1mm]
 \hspace*{-25mm}
G_{2,2}^{k,R}  &=&
 k(k-1)R^{-2}   +   2k(k-1)(k-2)R^{-3}
  +  k(k-1)(k-2)(k-3)R^{-4}
\end{eqnarray}

\end{document}